\documentclass[a4paper,UKenglish,cleveref, autoref, thm-restate]{lipics-v2021}

\pdfoutput=1 
\hideLIPIcs  
\nolinenumbers

\usepackage{todonotes}
\usepackage{tikz}
\usepackage{svg}

\usepackage{thmtools}
\usepackage{thm-restate}
\usepackage[T1]{fontenc}
\usepackage{xspace}
\usepackage[ruled,vlined,linesnumbered]{algorithm2e}

\newcommand{\eps}{\varepsilon}
\newcommand{\opt}{\mathrm{OPT}}

\newcommand{\twobp}{\textsc{2d-bp}\xspace}
\newcommand{\tbp}{\textsc{3d-bp}\xspace}
\newcommand{\tsp}{\textsc{3d-sp}\xspace}
\newcommand{\tmvc}{\textsc{3d-mvbb}\xspace}

\newcommand{\inter}{intermediate\xspace}
\newcommand{\bigy}{big\xspace}
\newcommand{\tin}{tiny\xspace}

\renewcommand{\epsilon}{\lambda}

\newtheorem{property}[theorem]{Property}

\newcommand{\area}{\mathrm{area}}

\title{Improved Approximation Algorithms for Three-Dimensional Bin Packing}

\author{Debajyoti Kar}{Department of Computer Science and Automation, Indian Institute of Science, India}{debajyotikar@iisc.ac.in}{https://orcid.org/0000-0003-4007-4648}{Debajyoti Kar is supported by Google PhD Fellowship and Microsoft Research India PhD Award.}
\author{Arindam Khan}{Department of Computer Science and Automation, Indian Institute of Science, India}{arindamkhan@iisc.ac.in}{https://orcid.org/0000-0001-7505-1687}{Arindam Khan's research is supported in part by Google India Research Award, SERB Core Research Grant
(CRG/2022/001176) on “Optimization under Intractability and Uncertainty”, Ittiam Systems CSR grant, and the Walmart
Center for Tech Excellence at IISc (CSR Grant WMGT-23-0001). }
\author{Malin Rau}{Department of Mathematical Sciences, Chalmers University of Technology and University of Gothenburg, SE-412 96 Göteborg Sweden}{malin.rau@chalmers.se}{https://orcid.org/0000-0002-5710-560X}{}

\authorrunning{D. Kar and A. Khan and M. Rau}

\Copyright{Debajyoti Kar and Arindam Khan and Malin Rau} 



\category{} 

\relatedversion{}

\begin{document}

\maketitle

\begin{abstract}
We study two fundamental three-dimensional (3D) geometric packing problems: 3D (Geometric) Bin Packing (\tbp), and 3D Minimum Volume Bounding Box (\tmvc), where given a set of  3D (rectangular) cuboids, the goal is to find an axis-aligned nonoverlapping packing of all cuboids.
In \tbp, we need to pack the given cuboids into the minimum number of unit cube bins. 
In \tmvc, the goal is to pack them into a cuboid box of minimum volume. 

It is NP-hard to even decide whether a set of rectangles can be packed into a unit square bin -- giving an (absolute) approximation hardness of 2 for \tbp. The previous best (absolute) approximation for both the problems follows from a result of Buchwald and Scheithauer (Int.~Trans.~Oper.~Res., 2016), yielding approximation ratios of $11$, and $5+\eps$, respectively, for \tbp and \tmvc. We provide improved approximation ratios of $6$, and $3+\eps$, respectively, for the two problems, for any constant $\eps > 0$.


For \tbp, in the asymptotic regime, Bansal, Correa, Kenyon, and Sviridenko  (Math.~Oper.~Res., 2006) showed that there is no asymptotic polynomial-time approximation scheme (APTAS) even when all items have the same height. Caprara (Math.~Oper.~Res., 2008) gave an asymptotic approximation ratio of $T_{\infty}^2 + \eps\approx 2.86$, where $T_{\infty}$ is the well-known Harmonic constant in Bin Packing. We provide an algorithm with an improved asymptotic approximation ratio of $3 T_{\infty}/2 +\eps \approx 2.54$.
Further, we show that unlike \tbp, \tmvc admits an APTAS.
\end{abstract}



\section{Introduction}
\label{sec:intro}
Three-dimensional (3D) packing problems are used to model several practical settings in production and transportation planning -- ranging from cargo management, manufacturing, 3D printing and prototyping, to cutting and loading applications. In the 1960s, Gilmore and Gomory \cite{gilmore1965multistage} introduced 3D packing in the context of the cutting stock problem in operations research, where given a stock material (3D cuboid), the goal is to cut out a set of required items (smaller 3D cuboids) by a sequence of end-to-end cuts. Around the same time, Meir and Moser \cite{meir1968packing} asked a combinatorial question: given a set of cubes, when can we pack them in a given cuboid? Since then, due to its inherent mathematical aesthetics, computational nature, and practical relevance, the study of 3D packing has led to the development of several techniques in mathematics, computer science, and operations research.

In this paper, we consider two classical 3D packing problems. 
In both of these problems, the input is a collection of (rectangular) cuboids (items), each specified by their height, width, and depth.
In the 3D Bin Packing (\tbp) problem, the goal is to output a packing of all the items using the minimum number of bins, where each bin is a unit cube. 
In the Minimum Volume Bounding Box (\tmvc) problem, we seek to obtain a cuboidal box of minimum volume that can accommodate all input items. 
In both these problems, the items cannot be rotated about any axis, and they must be packed non-overlappingly. Further, we assume that all items and bins/boxes are axis-aligned.

With the recent exponential growth in transportation and shipping, specially with the advent of e-commerce and UAVs, these problems are receiving increasingly more attention. 
For instance, in container ship loading, it is crucial to optimize the placement of cargo to maximize space usage while minimizing the number of containers needed. In pallet loading, manufacturers strive to stack goods on pallets in a way that maximizes storage capacity and ensures secure transport. Further, in supply chain management, it is crucial to optimize the arrangement of goods in storage to fit within the smallest possible space, reducing storage costs and enhancing inventory accessibility.
The survey by Ali, Ramos, Carravilla, and Oliveira~\cite{ali2022line} provides a comprehensive overview of 3D packing, 
with more than two hundred research articles. 
 We refer readers to \cite{faroe2003guided, hifi2010linear, lodi2002heuristic, crainic2009ts2pack, parreno2010hybrid, crainic2008extreme,jin2003three,li2014genetic, mahvash2018column,mack2012heuristic} 
 for important empirical procedures and heuristics to \tbp.
 There are also many practical programming competitions for \tbp, e.g., OPTIL 3D Bin Packing Challenge \cite{optil} and  ICRA VMAC Palletization  Competition  \cite{vmac}.

In contrast to the above, the theoretical exploration of 3D packing has been significantly limited due to its inherently complicated nature. 
Both the considered problems are NP-hard. In fact, \tbp generalizes several classical strongly NP-hard problems in scheduling and packing, including (1D) bin packing, multiprocessor scheduling \cite{li-cheng}, packing squares into squares \cite{ferreira1999packing}, and packing cubes into cubes \cite{lu2015packing}. 
In this paper, we study the absolute and asymptotic approximation algorithms for these problems. Given an algorithm $\mathcal{A}$ for a minimization problem $\Pi$, the {\em absolute approximation ratio} of $\mathcal{A}$ is defined as $\max_{I \in \mathcal{I}} \{\mathcal{A}(I)/\opt(I)\}$, where 
$\mathcal{I}$ is the set of all input instances for $\Pi$, and $\mathcal{A}(I), \opt(I)$ are the values of the solution provided by  $\mathcal{A}$ and the optimal solution for an  input instance $I$, respectively.
The {\em asymptotic approximation ratio} (AAR) is defined as: 
$\limsup\limits_{m\rightarrow \infty} \max_{I \in \mathcal{I}} \left\{\frac{\mathcal{A}(I)}{\opt(I)} \mid \opt(I)=m \right\}.$
A problem is said to admit an asymptotic polynomial-time approximation scheme (APTAS) if, for any $\eps>0$, there exists a polynomial-time algorithm $\mathcal{A}_\eps$ with AAR of $(1+\eps)$.
\tbp generalizes 2D Bin Packing. Thus it does not admit an APTAS, as 2D Bin Packing has an asymptotic approximation hardness of $1+1/2196$ \cite{chlebik2009hardness}. 
Furthermore, even for squares, it is NP-hard to decide if a set of squares can be packed in a single square bin or not \cite{ferreira1999packing} -- thus giving an absolute approximation hardness of 2 for \tbp.

Two-dimensional variants of these problems have been extensively studied. For \textsc{2d-bp}, Harren, Jansen, Pr{\"a}del, Schwarz, and van Stee~\cite{harren2013two} gave a {\em tight} absolute 2-approximation, and a line of work \cite{caprara2008packing,bansal2010new,jansen2016new} culminated in an asymptotic 1.406-approximation due to Bansal and Khan \cite{bansal2014improved}. 
For \textsc{2d-mvbb}, Bansal, Correa, Kenyon, and Sviridenko~\cite{bansal2006bin} gave a PTAS.

For \tbp, Csirik and van Vliet \cite{csirik1993line} gave an asymptotic approximation ratio of $T_{\infty}^3 + \eps \approx 4.836$, where $T_{\infty} \approx 1.691$ is the omnipresent  Harmonic constant \cite{LeeL85} in Bin Packing, and the same ratio was achieved by Epstein and van Stee \cite{epstein2004optimal} by an online algorithm using bounded space. 
This was later improved to $T_{\infty}^2 + \eps\approx 2.86$ by Caprara \cite{caprara2008packing}, which stands as the currently best-known asymptotic approximation ratio for \tbp. 

A related packing problem is 3D Strip Packing (\tsp), where we are given a three-dimensional strip having a $1\times 1$ square base and unbounded height, and we have to pack all items minimizing the height of the strip. 
For \tsp, Li and Cheng \cite{li-cheng} demonstrated that simple heuristics such as NFDH or FFDH for 2D packing \cite{coffman1980performance} have unbounded AARs. Then they provided an algorithm that returns a packing into a strip of height at most $(13/4)\opt_{\tsp}+8{h_{\max}}$, where $\opt_{\tsp}$ denotes the optimal Strip Packing height, and ${h_{\max}}$ is the maximum height of an input item. 
Afterwards, there has been a long line of work \cite{li-cheng,li1992heuristic,miyazawa2004packing,miyazawa1997algorithm,jansen2006asymptotic,bansal2007harmonic} on the asymptotic approximability of \tsp, culminating in a $(3/2+\varepsilon)$-approximation by Jansen and Pr{\"a}del \cite{3d-strip-packing}. However, all these improved asymptotic approximation algorithms incur huge additive loss (more than 100). 

The authors in \cite{3d-knapsack-diedrich} obtained an absolute approximation ratio of $(29/4+\varepsilon)$ for \tsp, and claimed it to be the best-known ratio for the problem in the abstract of their manuscript. However, in the main body of their manuscript, they presented an algorithm that returns a packing into a strip of height $4v(I)+2h_{\max}$. Since both $v(I)$ and $h_{\max}$ are lower bounds on $\opt_{\tsp}$, this already gives an absolute approximation ratio of $6$ for \tsp. 
Finally, Buchwald and Scheithauer \cite{buchwald2016upper} gave an absolute $5$-approximation for \tsp, which we believe to be the currently best-known absolute approximation ratio for the problem.
Since $\opt_{\tsp}$ is a lower bound on the minimum number of unit (cube) bins needed to pack all items, an absolute $\alpha$-approximation for \tsp~directly implies an absolute $(2\lfloor \alpha \rfloor +1)$-approximation for \tbp~-- one can obtain bins by cutting a \tsp solution at integral heights, followed by packing the items intersected by the cutting planes into additional separate bins. Thus, the $5$-approximation for \tsp due to \cite{buchwald2016upper} implies an absolute $11$-approximation for \tbp, which we believe to be the best-known approximation ratio for \tbp. 

Finally, it is easy to obtain an absolute $\alpha(1+\varepsilon)$-approximation for \tmvc from an absolute $\alpha$-approximation for \tsp.  Applying this strategy to the claimed $(29/4+\varepsilon)$-approximation algorithm for \tsp in the abstract of \cite{3d-knapsack-diedrich}, Alt and Scharf \cite{alt2018approximating} obtained a $(29/4+\varepsilon)$-absolute approximation for \tmvc. However, the result of Buchwald and Scheithauer \cite{buchwald2016upper} can also be extended to a $(5+\varepsilon)$-approximation for \tmvc.

There have been some improvements for special cases. 
For example, Bansal, Correa, Kenyon, and Sviridenko \cite{bansal2006bin} provided an APTAS for $d$-dimensional bin packing with $d$-dimensional cubes. 
Harren \cite{Harren09} gave APTAS for $d$-dimensional strip packing with $d$-dimensional cubes when the base of the strip has a bounded aspect ratio. 
Jansen, Khan, Lira, and Sreenivas \cite{Jansen0LS22} extended the APTAS to more general bases (not necessarily rectangular).

However, for general cuboids, there has been no progress on the absolute approximation ratios for both \tbp and \tmvc since 2016 \cite{buchwald2016upper} and for asymptotic approximation ratio of \tbp since 2008 \cite{caprara2008packing}.
In \cite{bansal2006bin}, the authors mention the inherent difficulty in extending results from 2D packing to 3D packing, due to the more complicated nature of interactions between different types of items in three dimension. 
In fact, improved approximability of $d$-dimensional geometric Bin Packing  for $d>2$, was listed as {\em one of the ten major open problems} in the survey on multidimensional packing \cite{christensen2016multidimensional}.

\subsection{Our contribution}

We present improved absolute and asymptotic approximation algorithms for \tbp, and \tmvc. 

First, we discuss our result on the absolute approximation algorithm for \tbp. 
We show how a packing in $k$ bins can be transformed into $6k$ {\em structured} bins; following which, for constant $k$, it is possible to find such a structured packing efficiently using a variant of the Generalized Assignment Problem -- giving us an absolute approximation ratio of 6.
One interesting idea is that we use an asymptotic approximation algorithm to obtain improved absolute approximation.
One of our key ingredients is the asymptotic approximation algorithm for \tsp by Jansen and Pr{\"{a}}del \cite{jansen2006asymptotic}, which provides a packing into height at most $(3/2+\varepsilon)\opt_{\tsp}+\varepsilon + O_{\varepsilon}(1)h_{\text{max}}$. \footnote{The notation $O_{\eps}(f(n))$ means that the implicit constant hidden in big-$O$ notation can depend on $\eps$.} 
Thus, for a sufficiently small appropriate constant $\mu$, if $h_{\text{max}}\le \mu$, then we can actually pack all items into $\lfloor 3k/2 \rfloor+1$ bins, assuming there exists a packing into $k$ bins. So, we partition the items into four classes: $L$ (large items: all dimensions are greater than $\mu$), $I_w, I_d, I_h$ (width, depth, height less than $\mu$, resp.).  If an item belongs to multiple classes, we assign them to anyone arbitrarily.
Now {\em large} items can be packed in $k$ bins by brute-force enumeration in polynomial-time (for constant $k,\mu$). 
Each of the remaining three classes can be packed into $\lfloor 3k/2 \rfloor+1$ bins. In total, we get $3(\lfloor 3k/2 \rfloor+1)+k\le 7k$ bins.

To improve further, we pack {\em large} items together with some items from one of the three classes. First, we observe that one of these classes has a volume less than $k/3$. W.l.o.g.~let us assume it to be  $I_h$. Now, first, we use a volume-based argument and use an algorithm from  \cite{li-cheng} to show that we can pack all items in $I_h$ whose width or depth is less than 1/2. The remaining items in $I_h$ have both width and depth exceeding 1/2. Next, we show that we can guess the packing of {\em large} items and almost all items in $I_h$, except a set of items with small volume. However, with a refined and technical analysis, we finally show that even these remaining items can be packed in the free regions of the six bins. For \tbp, this yields an improvement over the previous bound of 13 \cite{li-cheng}.

\begin{theorem}
\label{thm:3dbpabsolute}
    There exists a polynomial-time 6-approximation algorithm for \tbp.
\end{theorem}



This directly implies a $(6+\eps)$-approximation for the \tmvc~problem, using the connection between \tsp and \tmvc \cite{alt2018approximating}. 
However, we then use the power of resource augmentation in \twobp to obtain an APTAS for \tsp when we are allowed to use resource augmentation. With additional technical adaptations, we obtain a $(3+\eps)$-approximation for \tmvc.


Furthermore, surprisingly, unlike \tbp and \tsp, we show that \tmvc admits an APTAS -- settling the asymptotic approximability for the problem. 

\begin{theorem}
\label{thm:mvbb}
    For any $\varepsilon > 0$, there exists a polynomial-time $(3+\varepsilon)$-approximation algorithm and an asymptotic polynomial-time approximation scheme for \tmvc.
\end{theorem}

Finally, we turn our attention to the asymptotic approximability of \tbp.
Towards this, we exploit connections between \tsp and \tbp. 
Let $\opt_{\tsp}(I), \opt_{\tbp}(I)$ be the values of the optimal solution for Strip Packing and Bin Packing, for input $I$, respectively. Then $\opt_{\tsp}(I) \le \opt_{\tbp}(I)$, as the bins can be stacked on top of each other to provide a feasible solution for Strip Packing.
Thus, one can trivially obtain a $(3+\eps)$-approximation algorithm as follows. First, we obtain a packing in height $(\frac32+\eps)\opt_{\tsp}(I)+O_{\eps}(1)$ using \cite{3d-strip-packing}. 
We then can cut the strip into unit cube bins by cutting it at integral heights. All items that are completely contained within heights $[i, i+1)$ are packed in the $(2i+1)$-th bin. Remaining items that are cut by the $x$-$y$ axis-aligned plane at height $i$  (these items form one layer of items where each item has height at most $h_{\max} \le 1$) are packed in $(2i)$-th bin. This would give us a packing into $(3+\eps)\opt_{\tsp}(I)+O_{\varepsilon}(1)$ bins. 

To improve beyond $T_{\infty}^2$, our approach will be to find a packing such that the items that are cut do not have large heights. Towards this, we use {\em harmonic rounding}  \cite{LeeL85}, where the function $f_k$ rounds up $\alpha\in(1/k,1]$
to nearest larger number of the form $1/q$ where $q\in \mathbb{Z}$. Thus, for $\alpha_i\in(1/(q+1),1/q]$,  $f_k(\alpha_i):=1/q$, for $q \in [k-1]$.
Otherwise, $f_k(\alpha_i):=\alpha_i$.
It is well-known \cite{bansal2007harmonic} that, for any sequence $\alpha_1, \alpha_2, \dots, \alpha_n$, with $\alpha_i \in (0,1]$ and $\sum_{i=1}^n \alpha_i \le 1$, for a small enough $\eps$,
we have $\sum_{i=1}^n  f_{1/\eps}(\alpha_i) \le T_{\infty}+\eps \approx 1.691$. 

We first round the item heights in $I$ using $f_{1/\eps}$ to obtain a new set of items $I^{\infty}$ and obtain a 3D Strip Packing of them using the algorithm by \cite{jansen2006asymptotic}. Let $\opt_{\tbp}^{T_{\infty}}(I^{\infty})$ be the optimal number of $1\times 1 \times T_{\infty}$-sized bins needed to pack all items in $I^{\infty}$. 
  Then, it is easy to see that  $\opt_{\tsp}(I^{\infty}) \le T_{ \infty}\opt_{\tbp}^{T_{\infty}}(I^{\infty})$. \footnote{For simplicity, we are ignoring the $O_{\varepsilon}(1)$ in the following discussion in this section.} Then we have $\frac32\opt_{\tsp}(I^{\infty}) \le  \frac{3 T_{\infty}}{2}\opt_{\tbp}^{T_{\infty}}(I^{\infty}) \le \frac{3T_{\infty}}{2} \opt_{\tbp}(I)$. The last inequality follows from harmonic rounding. 
  
  Now we need to ensure that the {\em tall} items in $I^{\infty}$ packed in the strip with height $\frac32\opt_{\tsp}(I^{\infty})$ are not cut by the cutting planes at integral heights -- we call this {\em tall-not-sliced} property. A similar idea was used by Bansal, Han, Iwama, Sviridenko, and Zhang \cite{bansal2007harmonic} to obtain an alternate $(T_{\infty}+\eps)$-approximation for \twobp. However, 3D packing is much more involved.
  For this, we exploit the structural properties from the packing by \cite{3d-strip-packing}.
  First, we show that the strip can be divided into $O_\eps(1)$ cuboids such that, for each cuboid, the corresponding items packed inside are {\em similar}. 
  Next, we show that we can pack almost all {\em tall} items in $I^{\infty}$  of the same height ($1/q$ for some $q \in [k-1]$) at heights that are multiples of $1/q$  and incur only a small additive loss. This will ensure that none of these items are cut by planes at integral heights. For items with big width and depth, we use a linear program to assign items to the containers. For other items (except a small volume of them), the packing is based on variants of the Next-Fit-Decreasing-Height (NFDH) algorithm \cite{coffman1980performance}. 
  Finally, we show that we can pack the remaining items in the remaining free regions and an additional $O_{\eps}(1)$ bins. 
  This provides an improved guarantee for \tbp after nearly two decades.

\begin{theorem}
\label{thm:3dbpasymp}
    For any $\eps >0$, there exists a polynomial-time algorithm for \tbp~with an asymptotic approximation ratio $(3T_{\infty}/2+\eps) \approx 2.54$.
\end{theorem}

\begin{table}
    \centering
    \renewcommand{\arraystretch}{1.3} 
    \begin{tabular}{|c|>{\centering\arraybackslash}p{2.5cm}|>{\centering\arraybackslash}p{2.6cm}|
                      >{\centering\arraybackslash}p{2.6cm}|>{\centering\arraybackslash}p{2.5cm}|}  
        \hline
        \multirow{2}{*}{\textbf{Problem}} & \multicolumn{2}{c|}{\textbf{Absolute Approximation Ratio}} & \multicolumn{2}{c|}{\textbf{Asymptotic Approximation Ratio}} \\ 
        \cline{2-5}
        & \textbf{Previous Best} & \textbf{Our Result} & \textbf{Previous Best} & \textbf{Our Result} \\
        \hline
        \tbp & $11$ \cite{buchwald2016upper} & $6$ (WR: $5$) & $T_{\infty}^2+\varepsilon < 2.86$ \cite{caprara2008packing} & $3T_{\infty}/2+\varepsilon < 2.54$ \\
        \hline
        \tmvc & $5+\varepsilon$ \cite{buchwald2016upper, alt2018approximating} & $3+\eps$ & $5+\varepsilon$ \cite{buchwald2016upper, alt2018approximating} & $1+\varepsilon$ \\
        \hline
    \end{tabular}
    \caption{Summary of results. WR denotes the case when $90^{\circ}$ rotation around any axis is allowed. }
    \label{tab:comparison}
\end{table}

\noindent{\bf Organization of the paper.}
In Section \ref{sec:prelim}, we present some preliminaries needed for our results. 
Section \ref{sec:absolutesix} provides the absolute approximation algorithm for \tbp, and we prove Theorem \ref{thm:3dbpabsolute} 
Section \ref{sec:asymptotic} deals with the asymptotic approximation algorithm for \tbp and establishes \Cref{thm:3dbpasymp}. In Section \ref{sec:minvolumecontainer}, we discuss results related to \tmvc and prove \Cref{thm:mvbb}. 
Finally, Section \ref{sec:conc} ends with a conclusion. An overview of all results can be found in \cref{tab:comparison}.

\subsection{Related work}
For $d>3$, Caprara \cite{caprara2008packing} gave an algorithm with AAR of $T_{\infty}^{d-1}$ for both $d$-dimensional Bin Packing and Strip Packing. 
Sharma \cite{Sharma21}  gave $T_{\infty}^{d-1}$-asymptotic approximation for these two problems when the items
can be orthogonally rotated about all or a subset of axes.
For \tmvc, if the items are allowed to be rotated by 90 degrees about any axis, Alt and Scharf \cite{alt2018approximating} gave a 17.738-approximation.
Another related problem is the 3D Knapsack problem, in which each item additionally has an associated profit, and the goal is to obtain a maximum profit packing inside a unit cube knapsack. The authors in \cite{3d-knapsack-diedrich} have given a $(7+\varepsilon)$-approximation algorithm. For other related problems, we refer the readers to the surveys on approximation algorithms for multidimensional packing \cite{alt2016computational,christensen2016multidimensional}.


\section{Preliminaries}
\label{sec:prelim}
We define width, depth, and height along $x,y,z$ axes, respectively.  
Let $I$ be the given set of $n$ items, where each item $i\in I$ is an axis-aligned cuboid having height, width, and depth equal to $h_i, w_i, d_i$, 
respectively. Let $h_{\max}, w_{\max}, d_{\max} \in (0,1]$ be the maximum height, width, and depth of an item in $I$, respectively. 
Given a box $B:=[0,W]\times [0,D] \times [0,H]$, if the (bottom-left-back corner of) item $i$ is placed (by translation) at $(x_i, y_i, z_i)$ then it occupies the region: $[x_i, x_i+w_i] \times [y_i, y_i+d_i] \times [z_i, z_i+h_i]$, and the packing is feasible if $x_i \in [0,W-w_i],  y_i \in [0,D-d_i],  z_i \in [0,H-h_i]$. 
In this placement, we define the top, right, and back faces of item $i$ to be $[x_i,x_i+w_i]\times[y_i, y_i+d_i]\times\{z_i+h_i\}$, $\{x_i+w_i\}\times[y_i, y_i+d_i]\times[z_i, z_i+h_i]$, and $[x_i, x_i+w_i]\times\{y_i+d_i\}\times[z_i, z_i+h_i]$, respectively. Analogously, bottom, left, and front faces are defined.
Two items do not overlap if their interiors are disjoint. 
The volume of item $i$ is $v(i):=h_i w_i d_i$. 
For any set $T$, let $v(T)$ denote the total volume of items in $T$. We define $\opt_{\Pi}(I)$ to be the value of the optimal solution for problem $\Pi$ on instance $I$.

\subsection{Algorithms for 3D Packing}
We now state three results on 3D packing that will be crucial for our results.
The first two results give a volume-based guarantee.

\begin{restatable}[\cite{li-cheng}]{theorem}{licheng}
\label{thm:licheng}
    Let $T$ be a set of 3D items where each item has height bounded by $h_{\max}$. 
    \begin{enumerate}[(i)]
        \item All items in $T$ can be packed into a strip with $1\times 1$ base and height $4v(T) + 8h_{\max}$.
        \item If further, each item has either width or depth (or both) not exceeding $1/2$, then all items in $T$ can be packed inside a strip with $1\times 1$ base and height $3v(T)+8h_{\max}$.
    \end{enumerate}
\end{restatable}

\begin{proof}
    We first prove (ii). Let $T_w := \{i\in T \mid w_i \le 1/2\}$ and $T_d := T \setminus T_w$, so that $d_i \le 1/2$, for all $i\in T_d$. We further classify the items of $T_w$ based on their base area as follows: let $T_{w\ell}\subseteq T_w$ be the items whose base area exceeds $1/6$ and let $T_{ws} := T_w \setminus T_{w\ell}$. For packing items in $T_{w\ell}$, we first sort them in non-increasing order of heights and pack them in layers, with two items in each layer placed one beside the other (except possibly the last one). Note that this is possible since $w_i \le 1/2$ for $i\in T_w$. Let $k$ denote the number of layers and let $h'_1,\ldots, h'_k$ be the heights of the layers so that $h'_i$'s are non-increasing. Since each layer (except possibly the last) has a base area more than $1/3$, we have $v(T_{w\ell})> \frac{1}{3}\sum_{i=2}^{k} h_i'$, and therefore the height of the packing of $T_{w\ell}$ is at most $3v(T_{w\ell})+h_{\max}$.

    Next, we pack items of $T_{ws}$. To this end, we again sort the items in non-increasing order of heights and group them into maximal groups of base area at most $1/2$. Since the base area of each item is at most $1/6$, the total base area of each group (except possibly the last) is at least $1/2-1/6 = 1/3$. We pack each group in a layer using Steinberg's algorithm \cite{steinberg1997strip}. Similar to the packing of $T_{w\ell}$, the total height of the layers of $T_{ws}$ is bounded by $3v(T_{w\ell})+h_{\max}$.

    Analogously, we obtain a packing of items in $T_d$ in layers and place them above the layers of $T_w$. Altogether, we obtain a packing of $T$ inside a strip of height $3v(T)+4h_{\max}$.

    We now turn to the proof of (i). For this we classify $T$ as follows: let $T_{\ell} := \{i\in T \mid w_i > 1/2 \text{ and } d_i > 1/2\}$, and $T_s := T\setminus T_{\ell}$. We pack the items of $T_{\ell}$ in layers, with one item per layer. Since the base area of each item in $T_{\ell}$ exceeds $1/4$, the height of this packing is at most $4v(T_{\ell})$. For the set $T_s$, we use the packing algorithm of (ii) that yields a packing into a strip of height $3v(T_s)+4h_{\max}$. Placing this strip above the packing of $T_{\ell}$, we obtain a packing of $T$ inside a strip of height $4v(T)+4h_{\max}$, completing the proof. 
\end{proof}

\begin{theorem}[\cite{martello2000three}]
\label{thm:volume-packing}
    Given a set of 3D items $T$, there is a polynomial-time algorithm that places these items into at most $8v(T) + O(1)$ bins.
\end{theorem}

The last result is regarding the asymptotic approximation of \tsp. 

\begin{theorem}
[\cite{3d-strip-packing}]
\label{thm:jpstrippacking}
    Given a set of 3D items $I$ where each item has height bounded by $h_{\max}$, for any constant $\eps >0$, there is a polynomial-time algorithm that returns a packing of $I$ into a strip of height at most $(3/2+\varepsilon)\opt_{\tsp}(I)+\varepsilon + O_{\varepsilon}(1)h_{\max}$.
\end{theorem}

\subsection{Harmonic transformation}
\label{subsec:harmonic}
Lee and Lee \cite{LeeL85} introduced harmonic transformation in the context of online bin packing. 
The harmonic transformation with parameter $k$ is defined by the following function $f_k$:\\
For $\alpha_i\in(1/(q+1),1/q]$,  $f_k(\alpha_i):=1/q$, for $q \in [k-1]$.
Otherwise, $f_k(\alpha_i):=\frac{k}{k-1}\alpha_i$.\\
Intuitively, the function $f_k$ rounds up $\alpha_i\in(1/k,1]$
to the nearest larger number of the form $1/q$ where $q\in \mathbb{Z}$. 

Now we define harmonic constant $T_{\infty}$.
Let $t_1:=1$ and $t_{i+1}=t_i(t_i+1)$ for $i \in \mathbb{Z}_{\ge 2}$. 
The sequence $t_i+1$ is also known as Sylvester's sequence (where each term is the product of the previous terms, plus one).
Let $m(k)$ be the integer such that $t_{m(k)}\le k \le _{m(k)+1}$.
Now $T_k$ is defined as $\sum_{q=1}^{m(k)}\frac{1}{t_q}+\frac{k}{t_{(m(k)+1}\cdot (k-1)}$, and $T_{\infty}:=\lim_{k \rightarrow \infty} T_k$.
Thus $T_{\infty}=\sum_{i=1}^{\infty} \frac{1}{t_i}=1+\frac12+\frac16+\dots \approx 1.69103$.
 Note that $T_k \le T_{\infty}+\frac{1}{(k-1)}$.

Lee and Lee \cite{LeeL85} showed that that, for any sequence $\alpha_1, \alpha_2, \dots, \alpha_n$, with $\alpha_i \in (0,1]$ and $\sum_{i=1}^n \alpha_i \le 1$,
we have $ \sum_{i=1}^n  f_k(\alpha_i) \le T_{k}$.  
In fact, $\lim_{k \rightarrow \infty} \sum_{i=1}^n  f_k(\alpha_i) \le T_{\infty} \approx 1.691$. 

In fact, Bansal, Han, Iwama, Sviridenko, and Zhang \cite{bansal2007harmonic} showed the above inequality is true even if we define $f_k(\alpha_i):=\alpha_i$ for $\alpha_i \le 1/k$.

\subsection{Next-Fit-Decresing-Height (NFDH)}
\label{appx:NFDH}

The Next-Fit-Decreasing-Height (NFDH) algorithm is a shelf-based approach for packing 2D items into a strip of fixed width $w$. Given a set of items $I$, the algorithm first sorts them in decreasing order of height.
It then places the items sequentially from left to right on the floor of the strip (or the current shelf) until the next item no longer fits, i.e., adding the next item would cause the total width of items on the shelf to exceed $w$. At this point, a new shelf is created by drawing a horizontal line at the height of the tallest item on the current shelf. The process then continues on the new shelf, following the same placement rule, until all items have been packed.
For more details on the algorithm, we refer to \cite{coffman1980performance, christensen2016multidimensional}.

\begin{lemma}[\cite{coffman1980performance}]
\label{lem:NFDH}
Given a 2D rectangular box of height $h$ and width $w$, and a set of 2D items with maximum height $h_{\max}$ and maximum width $w_{\max}$, it is possible to place any subset of items with a total area of at most $(h-h_{\max})(w-w_{\max})$, into the box using NFDH.
\end{lemma}

\subsection{Generalized Assignment Problem (GAP)}
\label{sec:GAP}
In the Generalized Assignment Problem, we are given a set of $k$ knapsacks, each with an associated capacity $\{c_j\}_{j\in [k]}$, and a set of $n$ items, where each item $i\in [n]$ has size $s_{ij}$ and profit $p_{ij}$ for knapsack $j$. The goal is to obtain a maximum-profitable packing of a subset of the items into the knapsacks that respects the knapsack capacities, i.e., the total size of the items packed inside each knapsack does not exceed the capacity of the knapsack. For the general case of GAP, there is a tight $(\frac{e}{e-1}+\eps)$-approximation \cite{fleischer2011tight}. However for the special case when $k=O(1)$, there exists a PTAS.

\begin{theorem}[\cite{2dknapsack-lpacking}]
    For any $\eps > 0$, there is an algorithm for GAP with $k$ knapsacks running in $n^{O(k/\eps^2)}$ time that returns a packing of profit at least $(1-\eps)\opt$.
\end{theorem}



\section{Absolute 6-approximation for 3D-BP}
\label{sec:absolutesix}
In this section, our goal is to prove \Cref{thm:3dbpabsolute}. Let $K>0$ be a large constant such that the algorithm of Caprara \cite{caprara2008packing} already yields an absolute 6-approximation when $\opt_{\tbp}>K$. Our goal is to obtain a 6-approximation for the case when $\opt_{\tbp} \le K$. 
Let $\lambda = 1/40$, and $\delta<\lambda$ be a sufficiently small constant. 
The following lemma follows from a standard shifting argument.

\begin{restatable}{lemma}{medium}
\label{lem:medium}
    There exists a polynomial-time computable $\mu \le \delta$ such that the total volume of the items that have at least one of the dimensions in the range $(\mu^4,\mu]$ is at most $\delta$.
\end{restatable}

\begin{proof}
    Let $\mu_0 = \delta$, and for $j\in [3K/\delta]$, define $\mu_j = \mu_{j-1}^4$. Let $I_j \subseteq I$ be the items having at least one dimension in the range $(\mu_j, \mu_{j-1}]$. Note that since we are in the case when $\opt_{\tbp}\le K$, we have $v(I)\le K$. Observing that each item can belong to at most three of the sets $\{I_j\}_{j\in [3K/\delta]}$, we have $\sum_{j\in [3K/\delta]} v(I_j) \le 3K$, and thus there must exist an index $j^*$ for which $v(I_{j^*}) \le \delta$. Setting $\mu = \mu_{j^*-1}$ completes the proof.
\end{proof}

We classify the items depending on their dimensions: let $L$ be the items whose height, width and depth all exceed $\mu$ (called \emph{large} items), $I_h$ be the items with height at most $\mu^4$, $I_w$ be the remaining items having width at most $\mu^4$ and $I_d$ be the remaining items with depth at most $\mu^4$. Finally, let $I^{\text{rem}}$ be the remaining items, each having at least one of the dimensions in the range $(\mu^4,\mu]$. Note that $v(I^{\text{rem}})\le \delta$ owing to \Cref{lem:medium}. We further classify the items of $I^{\text{rem}}$ in a similar way -- let $I^{\text{rem}}_h \subseteq I^{\text{rem}}$ be the items with height at most $\mu$, $I^{\text{rem}}_w \subseteq I^{\text{rem}}\setminus I^{\text{rem}}_h$ be the remaining items with width at most $\mu$, and $I^{\text{rem}}_d = I^{\text{rem}} \setminus (I^{\text{rem}}_h \cup I^{\text{rem}}_w)$.

In the remainder of the section, we prove the following result.

\begin{proposition}
\label{pro:sixapprox}
    If there exists a packing of all items into $k\le K$ bins, then a packing using at most $6k$ bins can be computed in polynomial time.
\end{proposition}

\begin{figure}
    \centering
    \includegraphics[width=0.55\linewidth]{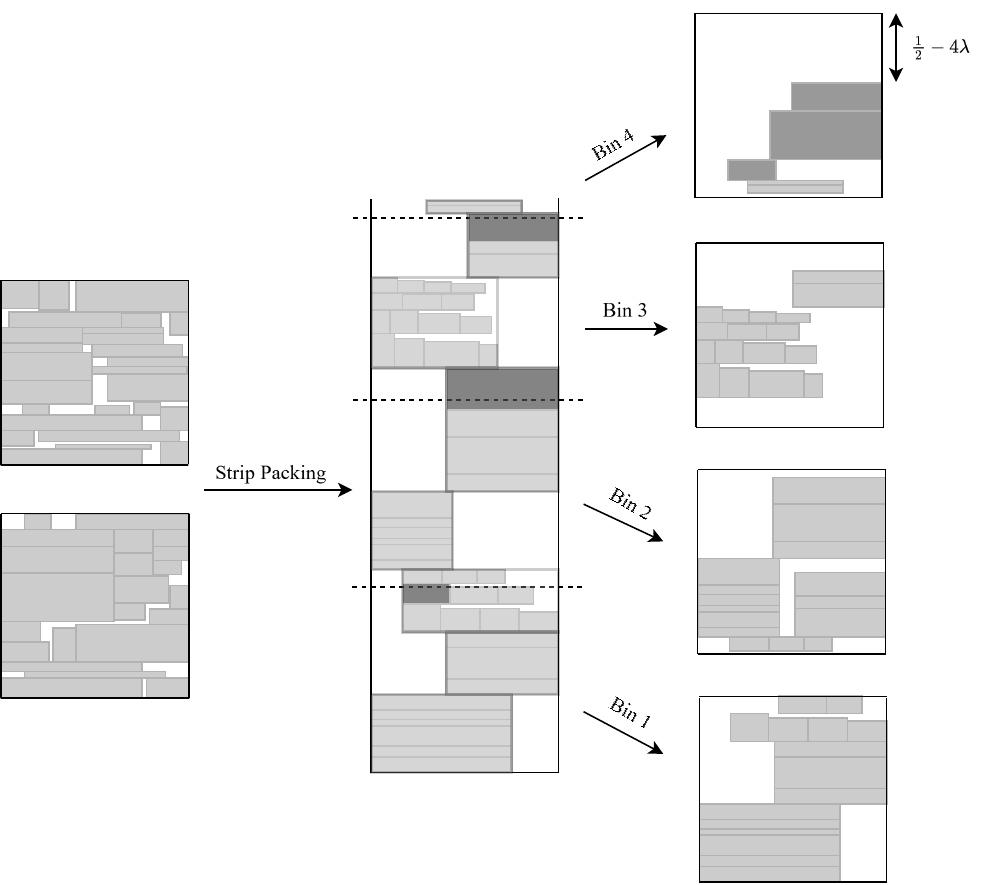}
    \caption{Packing from \Cref{thm:packseparate} (only the front view is shown for simplicity) for $k=2$. The dark gray items are sliced while cutting out $\lfloor 3k/2 \rfloor+1 $ bins from the Strip Packing solution. Finally, sliced items are packed into the empty regions of the last bin.}
    \label{fig:sptobp}
\end{figure}

For this, we first show the following lemma, which follows from a simple application of \Cref{thm:jpstrippacking}.

\begin{restatable}{lemma}{packseparate}
\label{thm:packseparate}
    Let $T$ be a set of items, each having a height (analogously width, depth) of at most $\mu$, and suppose that there exists a packing of $T$ into $k\le K$ bins. Then, it is possible to compute a packing of $T$ using $\lfloor 3k/2 \rfloor + 1$ bins in polynomial time. Further, one of these bins has an empty strip with $1\times 1$ base and height (analogously width, depth) $1/2-4\epsilon$. 
\end{restatable}

\begin{proof}
    Since there exists a packing of $T$ into $k$ bins, the optimal Strip Packing height of the items of $T$ is also bounded by $k$. Using \Cref{thm:jpstrippacking} with $\varepsilon = \epsilon/K$, we first obtain a packing of $T$ into a strip of height at most $(3/2+\epsilon/K)k + \epsilon/K + O_{\lambda}(1)\mu\le 3k/2 + 3\epsilon$, since $\mu \le \delta$ and $\delta$ is sufficiently small. Next, we cut the strip at integral heights and let $T'$ be the set of items that are sliced by this process. The remaining items of $T$ are thus packed into $\lceil 3k/2 + 3\epsilon \rceil = \lfloor 3k/2 \rfloor + 1$ bins. Note also that the last bin is filled up to a height of at most $1/2+3\epsilon$. We pack the items of $T'$ into the last bin, which further occupies a height of at most $\mu(\lfloor 3k/2 \rfloor + 1) \le \epsilon$. Hence, the empty region inside the last bin has a height of at least $1/2 - 4\epsilon$.
\end{proof}

We divide the proof of \Cref{pro:sixapprox} into two cases depending on $v(L)$.

\subsection{Case 1: \texorpdfstring{$v(L) > 64\delta K$}{large volume of L}}
In this case, for some $j\in \{h,w,d\}$, the total volume of the items in $I_{j}$ must not exceed $(k-64\delta K)/3 \le (1/3 - 21\delta)k$ -- w.l.o.g. assume that $j=h$. We first pack the items of $I_w \cup I^{\text{rem}}_w$ and $I_d \cup I^{\text{rem}}_d$ into $\lfloor 3k/2\rfloor +1$ bins each,  using \Cref{thm:packseparate}. Our goal next is to pack the items of $L \cup I_h \cup I^{\text{rem}}_h$ into $2k$ bins.
To this end, we further classify the items of $I_h$ depending on their width and depth. Let $I_{h,\ell} := \{i\in I_h \mid w_i, d_i > 1/2\}$ and let $I_{h,s}:= I_h \setminus I_{h,\ell}$. We first pack the items of $I_{h,s}$ into $k$ bins using the following lemma by applying \cref{thm:licheng}.

\begin{restatable}{lemma}{emptyregion}
\label{lem:emptyregion}
    The items of $I_{h,s}$ can be completely packed into $k$ bins where each bin additionally has an empty strip with $1\times 1$ base and height $59\delta$.
\end{restatable}
\begin{proof}
    We arbitrarily group the items of $I_{h,s}$ into maximal groups of volume at most $(1/3-20\delta)$ each. Since the volume of each item is bounded by $\mu^4$, each group has a volume of at least $1/3-20\delta-\mu^4 > 1/3-21\delta$, and therefore the number of groups is at most $k$ (since $v(I_{h,s})\le v(I_h)\le (1/3-21\delta)k$). Using \Cref{thm:licheng}, the items in each group can be packed in a bin within a height of $3\cdot (1/3-20\delta)+8\mu^4 \le 1-59\delta$, leaving an empty region of height $59\delta$ as claimed.
\end{proof}

Now consider the optimal packing inside $k$ bins restricted to the items of $I_{h,\ell}\cup L$. Our goal is to compute a packing of all these items, barring a subset of items from $I_{h,\ell}$ that have a total volume of at most $O(\mu)k$. 
For this, we first discretize the positions of the large items inside the bins. We focus on the packing inside any one bin.

\begin{figure}
    \centering
    \includegraphics[width=0.55\linewidth]{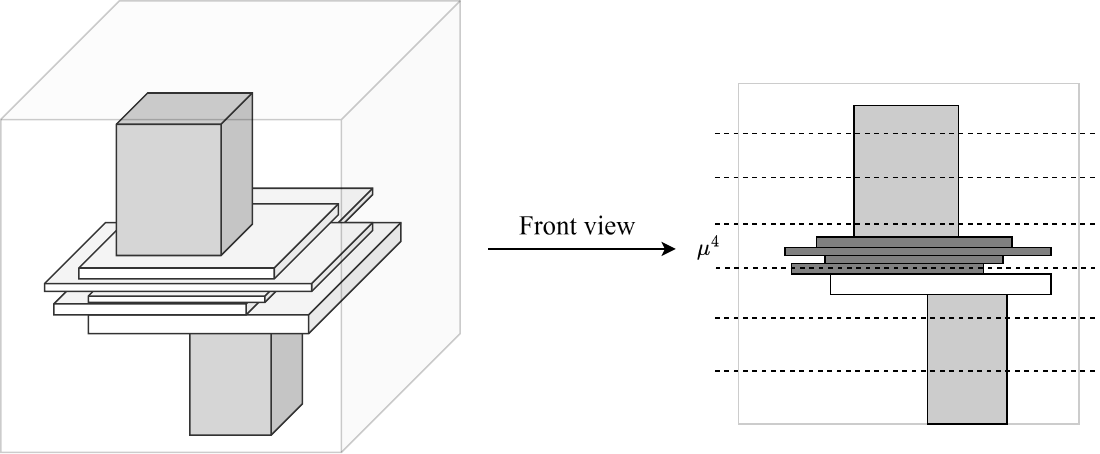}
    \caption{The light gray items are items of $L$. The dark gray items are deleted in order to position the upper large item at a multiple of $\mu^4$.}
    \label{fig:discretize}
    \end{figure}

\begin{lemma}
\label{lem:discretepos}
    By discarding items of $I_{h,\ell}$ having a total volume of at most $2\mu$, the number of distinct positions of the items of $L$ can be assumed to be polynomially-bounded. 
\end{lemma}
\begin{proof}
    Let us start with the optimal packing inside $k$ bins restricted to the items of $I_{h,\ell}\cup L$.
    For discretizing the $x$- and $y$-positions, we push all items as much to the left and front as possible. Then the distance of the left (resp.~front) face of an item of $L$ from the left (resp.~front) face of the bin can be written as the sum of the widths (resp.~depths) of at most $1/\mu$ large items and at most one item of $I_{h,\ell}$. Thus, there are polynomially many choices.

    In order to discretize the positions along the $z$-axis, we draw horizontal planes inside the bin at heights as integral multiples of $\mu^4$ and consider the large items inside the bin from bottom to top. For each $i \in L$ in the order of increasing height of their bases, we discard items of $I_{h,\ell}$ in order to pull the item $i$ downwards until the bottom face of $i$ hits one of the horizontal planes at heights multiples of $\mu^4$ or the top face of another large item lying below $i$ (see \Cref{fig:discretize}). At the end of this process, it can be seen that the distance between the bottom face of any large item and the base of the bin can be written as the sum of a multiple of $\mu^4$ and the heights of at most $1/\mu$ items of $L$. The total volume of the discarded items is bounded by $(1/\mu^3)\cdot 2\mu^4 = 2\mu$.     
\end{proof}

\begin{figure}
    \centering
    \includegraphics[width=0.55\linewidth]{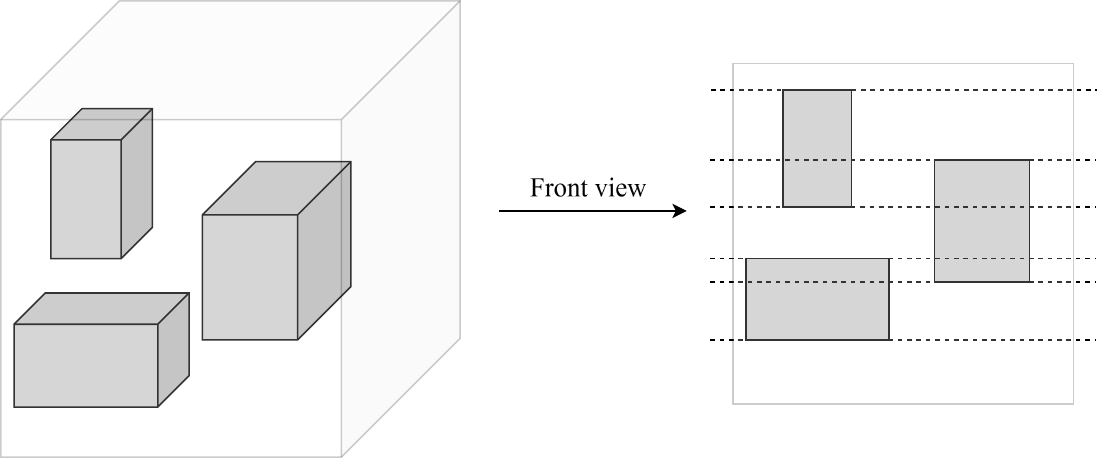}
    \caption{The regions between two consecutive dotted lines correspond to slots.}
    \label{fig:slots}
\end{figure}

We next draw horizontal planes passing through the top and bottom faces of each large item and discard the items of $I_{h,\ell}$ that are intersected by these planes. The volume of these discarded items is bounded by $(2/\mu^3)\cdot \mu^4 = 2\mu$. This partitions the bin into at most $2/\mu^3 +1$ \emph{slots}, where each slot is penetrated from top to bottom by at most $1/\mu^2$ large items (see \Cref{fig:slots}). Note also that no large item begins or ends in the interior of a slot. Together with \Cref{lem:discretepos}, we thus have the following result.

\begin{restatable}{lemma}{formslots}
\label{lem:formslots}
    There exists a subset $I'_{h,\ell}\subseteq I_{h,\ell}$ with $v(I'_{h,\ell})\ge v(I_{h,\ell})-4\mu k$ such that the items of $I'_{h,\ell}$ are completely packed inside the slots formed by the large items in the $k$ bins.
\end{restatable}
\begin{proof}
    By \Cref{lem:discretepos}, we discard items of $I_{h,\ell}$ having a volume of $2\mu k$, in order to discretize the positions of the large items. Also, the volume of items discarded while partitioning the $k$ bins into slots is bounded by $2\mu k$. The remaining items of $I_{h,\ell}$, that have a volume of at least $v(I_{h,\ell})-4\mu k$ are packed inside the slots in the $k$ bins.
\end{proof}

Our algorithm essentially tries to compute a packing close to the one guaranteed by the above lemma. As mentioned before, we obtain a packing into $k$ bins of all items of $L$, and a large volume subset of $I_{h,\ell}$ that is packed inside the slots formed by the large items. 

\begin{lemma}
\label{lem:GAP}
    In polynomial-time, it is possible to compute a set $I''_{h,\ell} \subseteq I_{h,\ell}$ with $v(I''_{h,\ell})\ge v(I_{h,\ell})-5\delta k$, and a packing of all items in $I''_{h,\ell}\cup L$ into $k$ bins.
\end{lemma}
\begin{proof}
    We first guess the positions (from polynomially many choices) of all the (at most $K/\mu^3$) large items inside the $k$ bins, and create slots by extending their top and bottom faces. Our goal is to pack a maximum volume subset of $I_{h,\ell}$ into these slots by creating an instance of the Generalized Assignment Problem (GAP) with $O(1)$ knapsacks, for which there exists a PTAS \cite{2dknapsack-lpacking} (see \Cref{sec:GAP} for details). Each slot corresponds to a knapsack with a capacity equal to the height of the slot. For an item $i\in I_{h,\ell}$, the size of $i$ for a slot equals its height $h_i$ if the item fits inside the slot, given the relative positions of the large items inside it, and $\infty$ otherwise. The profit of an item is the same as its volume. 
    By \Cref{lem:formslots}, the optimal packing packs items of $I_{h,\ell}$ having a volume of at least $v(I_{h,\ell})-4\mu k$. We apply the PTAS for GAP with parameter $\delta/K$ to our instance, yielding a packing of a subset $I''_{h,\ell} \subseteq I_{h,\ell}$ into the slots formed by the items of $L$, whose volume is least $(1-\delta/K)(v(I_{h,\ell})-4\mu k) \ge v(I_{h,\ell})-5\delta k$.
\end{proof}

It remains to pack the items of $I_{h,\ell}\setminus I''_{h,\ell}$ and $I^{\text{rem}}_h$. Intuitively, they have a small volume, and hence, we can pack them into the empty regions inside the already-existing bins.

\begin{restatable}{lemma}{repackdiscarded}
\label{lem:repackdiscarded}
    The items of $I_{h,\ell}\setminus I''_{h,\ell}$ can be completely packed by using a height of $25\delta$ from each of the empty regions inside the bins that were used to pack the items of $I_{h,s}$. Further, the items in $I^{\text{rem}}_h$ can be packed within a height of $12\delta$ inside one of the bins.
\end{restatable}
\begin{proof}
    Recall that each of the $k$ bins used for packing items of $I_{h,s}$ had an empty strip of height $59\delta$ by \Cref{lem:emptyregion}. Consider the items of $I_{h,\ell}\setminus I''_{h,\ell}$ that have a volume of at most $5\delta k$ by \Cref{lem:GAP}. We arbitrarily group them into maximal groups of volume not exceeding $6\delta$ each. Since each item in $I_{h,\ell}$ has a volume of at most $\mu^4$, the volume of each group is at least $6\delta - \mu^4 > 5\delta$, and thus there are at most $k$ groups. Using \Cref{thm:licheng}, the items in each such group can be packed within a height of $4\cdot 6\delta + 8\mu^4 \le 25\delta$.

    Note that there is still an empty region of height $59\delta - 25\delta = 34\delta$ left in each bin. 
    Since $v(I^{\text{rem}}_h)\le \delta$ by \Cref{lem:medium}, the items of $I^{\text{rem}}_h$ can be packed such that the packing has a height of at most $4\delta + 8\mu \le 12\delta$ using \Cref{thm:licheng}, and hence we can place them inside the remaining empty region inside any of these bins.
\end{proof}

\begin{figure}
    \centering
    \includegraphics[width=0.5\linewidth]{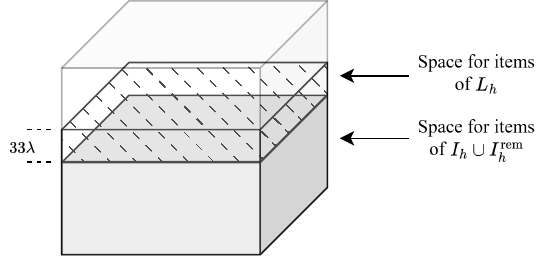}
    \caption{Packing inside bin $B_h$}
    \label{fig:Planes cutting}
\end{figure}

Altogether, we used $2\cdot (\lfloor 3k/2\rfloor +1)\le 4k$ bins for packing items in $I_w \cup I^{\text{rem}}_w \cup I_d \cup I^{\text{rem}}_d$ and $2k$ bins for packing the items of $L\cup I_h \cup I^{\text{rem}}_h$, resulting in at most $6k$ bins overall.

\subsection{Case 2: \texorpdfstring{$v(L)\le 64\delta K$}{small volume for L}}
\label{sec:largehassmallvol}

In this case, we first pack the items of $I_h\cup I^{\text{rem}}_h, I_w \cup I^{\text{rem}}_w$ and $I_d \cup I^{\text{rem}}_d$ into $\lfloor 3k/2 \rfloor + 1$ bins each, using \Cref{thm:packseparate} (note that these items have height, width and depth bounded by $\mu$, respectively). Let $B_h, B_w, B_d$ be the bins having empty strips of height, width, and depth $1/2-4\epsilon$, respectively, that are guaranteed by \Cref{thm:packseparate}. Intuitively, since the large items have a very small volume, they can be completely packed inside these empty strips.

\begin{restatable}{lemma}{packlarge}
\label{lem:packlarge}
    The items in $L$ can be completely packed inside three strips, each having a $1\times 1$ base aligned with the $xy$-, $yz$-, and $zx$-planes, respectively. The strips have height, width, and depth of $33\epsilon$, respectively, and therefore they fit inside the bins $B_h, B_w,$ and $B_d$.
\end{restatable}
\begin{proof}
    First, note that for sufficiently small $\delta$, we have $\delta K \le \epsilon^3$, and since $v(L) \le 64\delta K \le 64\lambda^3$, it follows that one of the dimensions of each large item must be at most $(64\epsilon^3)^{1/3}=4\epsilon$. We classify the items of $L$ into three groups -- let $L_h \subseteq L$ be those items with height at most $4\epsilon$, $L_w \subseteq L\setminus L_1$ be the items with width at most $4\epsilon$ and $L_d = L\setminus (L_1\cup L_2)$ be the remaining items, each having depth not exceeding $4\epsilon$. 
    
    Next, since $v(L_h)\le v(L)\le 64\lambda^3$, \Cref{thm:licheng} implies that the items in $L_h$ can be completely packed within a strip with $1\times 1$ base and a height of at most $4\cdot 64\lambda^3 + 8\cdot 4\epsilon \le 33\epsilon$ (see \Cref{fig:Planes cutting}). Analogously, we pack the items of $L_w$ and $L_d$ into strips of width and depth $33\lambda$, respectively.
\end{proof}

\noindent Overall, we obtain a packing into $3\cdot (\lfloor 3k/2 \rfloor +1) \le 6k$ bins, establishing \Cref{pro:sixapprox}.

\textbf{Overall algorithm:} We first run the algorithm of Caprara \cite{caprara2008packing} that already returns a 6-approximate solution when $\opt_{\tbp} > K$. Next, for each guessed value of $\opt_{\tbp}=k \le K$, we run the algorithm of \Cref{pro:sixapprox}. For this, we first compute a value of $\mu$ using \Cref{lem:medium} and classify the items as discussed.
Next, we divide into two cases depending on the volume of the large items. If $v(L) > 64\delta K$, we find $j \in \{h,w,d\}$ for which the volume of the items in $I_j$ does not exceed $(1/3-21\delta)k$; w.l.o.g~we take $j=h$. We pack the items of $I_w \cup I^{\text{rem}}_w$ and $I_d \cup I^{\text{rem}}_d$ into $\lfloor 3k/2\rfloor +1$ bins each using \Cref{thm:packseparate}. We classify items of $I_h$ into $I_{h,\ell}$ and $I_{h,s}$ depending on their width and depth and obtain a packing of $I_{h,s}$ into $k$ bins, ensuring each of these bins has an empty strip of height $59\delta$. Next we compute a set $I''_{h,\ell}\subseteq I_{h,\ell}$ such that $v(I_{h,\ell}\setminus I''_{h,\ell}) \le 5\delta k$, and pack the items of $I''_{h,\ell}\cup L$ into $k$ bins via a reduction to the Generalized Assignment Problem (\Cref{lem:GAP}). Finally, the items in $(I_{h,\ell}\setminus I''_{h,\ell})\cup I^{\text{rem}}_h$ are packed into the empty spaces inside the bins for $I_{h,s}$ using \Cref{lem:repackdiscarded}. For the other case when $v(L)\le 64\delta K$, we pack items in $I_h\cup I^{\text{rem}}_h, I_w \cup I^{\text{rem}}_w$ and $I_d \cup I^{\text{rem}}_d$ into $\lfloor 3k/2 \rfloor + 1$ bins each, using \Cref{thm:packseparate}, ensuring one of the bins has a sufficiently large empty strip, and then pack items of $L$ inside these empty strips using \Cref{lem:packlarge}.

\section{Asymptotic \texorpdfstring{$(\frac{3}{2}\cdot T_{\infty}+\eps)$}{near three-half}-approximation for 3D-BP}
\label{sec:asymptotic}

In this section, we prove \Cref{thm:3dbpasymp} and present an improved asymptotic approximation algorithm for \tbp.  
We will utilize ideas from the algorithm for \tsp by Jansen and Prädel~\cite{3d-strip-packing} which  packs $I$ into strip height of $(3/2+\varepsilon) \cdot \opt_{\tsp}(I) + O_{\eps}(1)h_{\max}$.
As mentioned earlier, the naive approach of cutting the strip at integral height will result in  $(3+\eps)$-approximation. 
Instead, we exploit the structural properties of the solution provided by the algorithm, along with the {\em harmonically rounded heights} of the items, to ensure that items with a height larger than $\eps$ are not sliced.

Recall the definition of harmonic rounding: 
For $\alpha_i\in(1/(q+1),1/q]$, $f_k(\alpha_i):=1/q$, for $q \in [k-1]$; and for $\alpha_i \in(0,1/k], f_k(\alpha_i):=\alpha_i$.
Also, if $\sum_{i=1}^n \alpha_i \le 1$,
then $\lim_{k \rightarrow \infty} \sum_{i=1}^n  f_k(\alpha_i) \le T_{\infty} \approx 1.691$. In the following, we assume $k=1/\eps$ to be large enough such that $\sum_{i=1}^n  f_{1/\eps}(\alpha_i)\approx T_{\infty}$, and define $f_{1/\eps}$ to be $f$. 

Let \( I^{\infty} \) be the instance derived from the given \tbp instance \( I \) by applying harmonic rounding $f$ to the heights of the items.
Thus an item $i \in I$ becomes an item of  $I^{\infty}$ with width, depth, height to be $w_i, d_i, f(h_i)$, respectively. 
Let \( \opt_{\tbp}^{T_\infty}(I^{\infty}) \) denote the minimum number of bins with dimensions \( 1 \times 1 \times T_{\infty} \) required to pack all items from \( I^{\infty} \), and let \( \opt_{\tsp}(I^{\infty}) \) denote the minimum height to pack all items from \( I^{\infty} \) into a strip with unit square base and unbounded height. 
The following lemma connects packing of $I^{\infty}$ with $I$:

\begin{restatable}{lemma}{asympHarmonic}
\label{lem:asymp-harmonic}
\( \opt_{\tsp}(I^{\infty}) \leq T_{\infty} \cdot  \opt_{\tbp}^{T_\infty}(I^{\infty}) \leq T_{\infty}\cdot \opt_{\tbp}(I) \).
\end{restatable}

\begin{proof}
First, we show that \( \opt_{\tbp}^\infty(I^{\infty}) \leq \opt_{\tbp}(I) \).
Take the optimal solution for $\tbp$, then increase the item heights due to the harmonic rounding and extend the height of the bin by $T_{\infty}$.
As the height of each combination of items that are on top of each other in the solution $\opt_{\tbp}(I)$ adds up to at most $1$, the sum of rounded heights will be bounded by  $T_{\infty}$, end hence the packing still fits into the bin.

Next, we prove that \( \opt_{\tsp}(I^{\infty}) \leq T_{\infty} \cdot  \opt_{\tbp}^{T_\infty}(I^{\infty}) \).
Consider the optimal solution to corresponding to $\opt_{\tbp}^{T_\infty}(I^{\infty})$.
We can create a solution to $\tsp$ by stacking all the bins on top of each other. 
As each bin has a height of at most $T_{\infty}$, the total height of the packing is bounded by $T_{\infty} \cdot  \opt_{\tbp}^{T_\infty}(I^{\infty})$ and hence
\( \opt_{\tsp}(I^{\infty}) \leq T_{\infty} \cdot  \opt_{\tbp}^{T_\infty}(I^{\infty}) \).
\end{proof}

We transform the instance $I^{\infty}$ for $\tsp$ into an instance for \twobp, similar to \cite{3d-strip-packing}, which then uses structural results from a {\twobp} algorithm \cite{jansen2016new}.
Given $\eps$, we round up the item heights of $I^{\infty}$  to the next multiple of $\frac{\eps v(I^{\infty})}{n}$.
We create an instance $I_{\twobp}$ of \twobp, by introducing for each item $i \in I^{\infty}$ with rounded height $k \frac{\eps v(I^{\infty})}{n}$, exactly $k$ rectangles with width $w_i$ and depth $d_i$. The following lemma bounds the incurred loss. 

\begin{restatable}{lemma}{asympTspToTwobsp}
\label{lem:asymp-tsp-to-twobsp}
$\frac{\eps v(I^{\infty})}{ n} \opt_{\twobp}(I_{\twobp})\leq (1+\eps)\opt_{\tsp}(I^{\infty})$.
\end{restatable}

\begin{proof}
We prove that $\frac{\eps v(I^{\infty})}{ n} \opt_{\twobp}(I_{\twobp})\leq (1+\eps)\opt_{\tsp}(I^{\infty})$.
    First, note that by rounding the item heights to the next larger multiple of $\frac{\eps v(I^{\infty})}{n}$ the total packing height of the optimal solution for \tsp is increased by at most $n \cdot \frac{\eps v(I^{\infty})}{n} = \eps v(I^{\infty})$, as there are at most $n$ items on top of each other.
    As each packing for \tsp given $I_{\tsp}$ has a height of at least $v(I^{\infty})$, the height of the instance is increased by at most $\eps v(I^{\infty}) \leq \eps\opt_{\tsp}(I^{\infty})$.

    Note that we can assume that in an optimal solution, each item either stands on the bottom or directly on top of another item. 
    We cut the optimal solution for the rounded instance at multiples of $\frac{\eps v(I^{\infty})}{n}$.
    Using these cuts, there is only one item on top of each other.
    Therefore, these slices represent a solution for $I_{\twobp}$ and hence $\frac{\eps v(I^{\infty})}{n} \opt_{\twobp}(I_{\twobp})\leq (1+\eps)\opt_{\tsp}(I^{\infty})$.
\end{proof}

\subsection{Essentials of the 2D-BP algorithm}
\label{subsec:2dbp}

The main ingredient for the \twobp~algorithm by Jansen and Prädel~\cite{jansen2016new} is a restructuring theorem.
It states that each packing can be rearranged into packing with an iterable structure. This rearrangement comes at the cost of introducing more bins to the packing.

First, the items are classified into \bigy, vertical, horizontal, \inter, and  \tin based on a suitably chosen constant $\delta \in [\eps^{O_{\eps}(1)}, \eps]$.
Let $\mu = \delta^4$.
An item $i$ is {\em \bigy} if $w_i \geq \delta$ and $d_i \geq \delta$; {\em vertical} if $d_i \geq \delta$ but $w_i < \mu$; {\em horizontal} if $w_i \geq \delta$ but $d_i < \mu$; {\em \tin} if $d_i < \mu$ and $w_i < \mu$; and {\em \inter} if either $w_i \in [\mu,\delta)$ or $d_i \in [\mu,\delta)$.
Using standard argument (see \Cref{lem:delta-twobp} in \Cref{subsec:2dbpAppndix}), we show that the area of \inter items is at most $\eps \cdot \area(I)$.

Jansen and Prädel~\cite{jansen2016new} showed the existence of a structured packing that we call \emph{$k$-2D-container-packing}.
In a \emph{$k$-2D-container-packing}, one can consider a rounded-up instance $\tilde{I}$ from $I_{\twobp}$, where the widths and depths of \bigy items, widths of horizontal items, and depths of vertical items from $I_{\twobp}$ are rounded up to $O(1/\delta^2)$ values.
Let $\mathcal{T}, \mathcal{W}, \mathcal{D}$ be the set of different types of (rounded) large items, widths of wide items, and depths of vertical items in $\tilde{I}$, respectively. 
Then $|\mathcal{T}|$ is $O(1/\delta^4)$, and $|\mathcal{W}|, |\mathcal{D}|$ are $O(1/\delta^2)$.

In a \emph{$k$-2D-container-packing}, each bin of the packing is partitioned into containers. 
Furthermore, vertical items are allowed to be sliced along $y$-axis, horizontal items may be sliced along $x$-axis, and \tin items may be sliced in both directions.
The containers are of five types, and only specific types of items from $\tilde{I}$ are allowed to be packed in the corresponding containers:
\begin{itemize}
\item (i) {\em Big containers:} Each such container contains only one (rounded up) \bigy item and has the size of this \bigy item. 
\item (ii) {\em Horizontal containers:} Each such container has a width $w \in \mathcal{W}$ and a depth that is a multiple of $\mu$, and contains only horizontal items with width $w$.  
Per bin, the total width of these containers is bounded by $O(k)$.
\item (iii) {\em Vertical containers:} 
Each such container has a depth $d \in \mathcal{D}$, a height that is a multiple of $\mu$, and contains only vertical items with depth $d$.
Per bin, the total depth of these containers is bounded by $O(k)$.
\item (iv) {\em Tiny containers:} Each such container contains only \tin items and has a width and depth that is a multiple of $\mu$. Each bin contains at most $O(k)$ of these containers.
\item (v) {\em Intermediate containers:} These containers contain only \inter items. There will be an extra $O(\eps)\opt_{\twobp}$ bins reserved separately to pack the \inter items.
\end{itemize}
We refer to a bin configuration as a valid way to split one bin into containers with regard to the considered $k$-2D-container-packing.
Jansen and Prädel~\cite{jansen2016new} showed the existence of a good {\em 2D-container-packing} (see \Cref{subsec:2dbpAppndix} for omitted details). 

\begin{restatable}[\cite{jansen2016new}]{theorem}{twoBPStructure}
\label{thm:finding-2D-container-packing}
    Given some instance $I$ for \twobp, there is an algorithm that finds in polynomial time a rounded-up instance $\tilde{I}$ with $|\mathcal{T}|, |\mathcal{D}|, |\mathcal{W}| \in O(1/\delta^4)$ and a $(1/\delta^3)$-2D-container-packing of $\tilde{I}$ into $B$ bins with at most $O_{\eps}(1)$ different bin-configurations that fulfills $B \leq (\frac{3}{2}+O(\eps))\opt_{\twobp}(I) +O_{\eps}(1)$.
\end{restatable}
Although this theorem is not explicitly stated in~\cite{jansen2016new}, it encapsulates the core idea of the paper. For completeness, we provide a proof in~\cref{subsec:2dbpAppndix}.

\subsection{The algorithm for 3D-BP}
\label{subsec:3dbpasymp}
\begin{algorithm}[t]
\DontPrintSemicolon
\caption{Asymptotic $((3/2)T_{\infty}+\eps)$-approximation for 3D-BP with input $I$ and $\eps>0$.}
\label{alg:asymp-tbp}
Create $I^\infty$ by harmonically rounding the item heights of $I$.\;
Create $I_{\twobp}$ after rounding the heights of the items in $I^\infty$ to multiples of $\frac{\eps v(I^{\infty})}{ n}$.\;
Find a rounded instance $\tilde{I}$ and the $(1/\delta^3)$-2D-container-packing into $B$ 2D bins with at most $O_{\eps}(1)$ bin-configurations.\;
Place $I^\infty$ into 3D containers corresponding to the \twobp solution as discussed in \cref{subsec:3dbpasymp}.\;
\textbf{return} the packing. \;
\end{algorithm} 

In this subsection, we describe how the items are filled into the bins when a $(\eps/\mu)$-2D-container-packing is given.
An overview of the complete algorithm can be found in \cref{alg:asymp-tbp}.
To prove the following Lemma, we consider the \bigy, horizontal and vertical, as well as \tin items separately.

\begin{restatable}{lemma}{algorithmToPlaceItemsIntoConfigurations}
\label{lem:alg-to-put-all-items-into-2D-container-packing}
    Given a rounded \twobp instance $\tilde{I}$ derived from the \tbp instance $I^{\infty}$ and an $(\eps/\mu)$-2D-container-packing of $\tilde{I}$ into $B$ 2D bins using $k$ 2D bin-configurations, a packing of the items $I^{\infty}$ into $(1+O(\eps))\frac{\eps v(I^{\infty})}{n} B + O(\log(1/\eps)k +|\mathcal{T}|)$ 3D bins can be found in polynomial time. 
\end{restatable}

\begin{figure}
    \centering
    \includegraphics[width=0.9\linewidth]{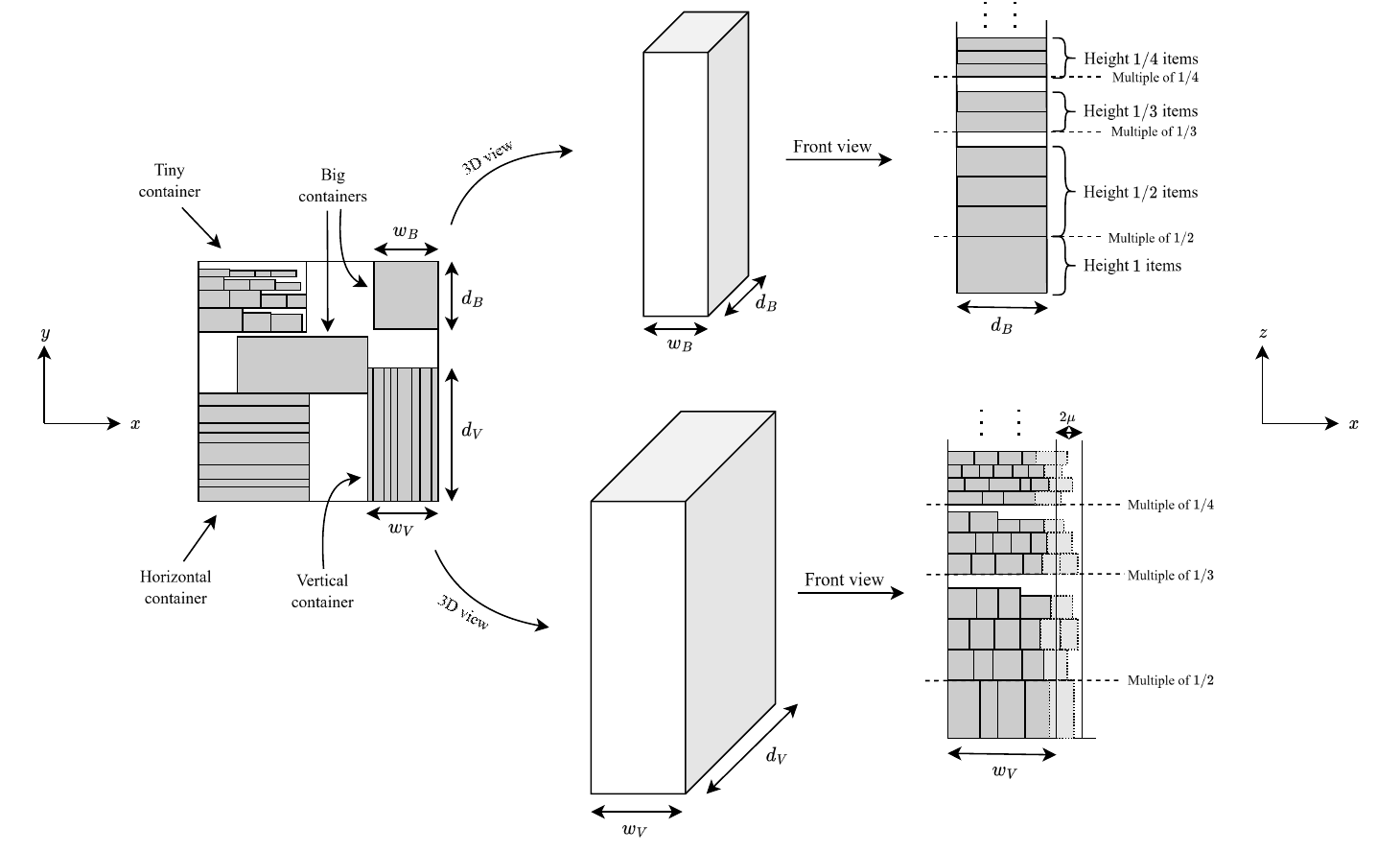}
    \caption{The left figure depicts a 2D container-packing, which forms the base of a 3D configuration. 
    The middle figure shows two 3D containers corresponding to two containers (one big and one horizontal) in 2D container-packing. 
    On the right, the packing that ensures the \emph{tall-not-sliced} property is shown. The light gray rectangles are repacked into additional bins.}
    \label{fig:container}
\end{figure}

Let $C$ be one of the $k$ 2D bin configurations used, and let $x_C$ denote the multiplicity (possibly fractional) of $C$ in the solution consisting of $B$ 2D bins.
We create a 3D configuration of height $\lceil x_C \cdot \frac{\eps v(I^{\infty})}{n}\rceil$, whose base is identical to $C$. 
Note that the rounding to an integer height increases the total height of the configurations by at most $k$, i.e., the total height of all 3D configurations is bounded by $\frac{\eps v(I^{\infty})}{n}B+k$.
The 2D containers of $C$ raised by the corresponding height of the configuration will be referred to as 3D containers and will be denoted by the same type as the corresponding 2D containers.

Our goal is to pack the items of $I^{\infty}$ into these 3D containers while ensuring the \emph{tall-not-sliced} property. Items with a rounded height larger than $\eps$ are called \emph{tall} items, and the remaining items are called \emph{short}. The classification of items into big, vertical, horizontal, tiny, and intermediate, depending on the dimensions of their top faces, as defined in the previous subsection, continues to hold.

{\bf Placing big items:} 
The different slices of a \bigy 3D item might be rounded differently by the \twobp~algorithm. 
However, using a linear program, we can assign all big 3D items except for $O(|\mathcal{T}|)$ to the big 3D containers without violating the container heights.
The $O_{\eps}(1)$ unassigned big items will be packed using $O_{\eps}(1)$ additional bins. 
To formalize this process,
consider the set of 3D items $I_B$ whose slices are \bigy 2D items in $\tilde{I}$.
Further, for any $t \in \mathcal{T}$, denote by $K_t$ the set of 3D containers that are designated for \bigy items with rounded size of type $t$ and let $\tilde{I}_t$ denote the set of 3D items with rounded size $t$.

\begin{lemma}
\label{lem:frist-setp-big-items}
Given a rounded (2D) instance $\tilde{I}$ of $I_{\twobp}$,
in polynomial time, it is possible to compute an assignment of all but $|\mathcal{T}|$ \bigy 3D items to rounded sizes $t \in \mathcal{T}$, such that for each $t \in \mathcal{T}$, the total height of assigned 3D items is bounded by $\frac{\eps v(I^{\infty})}{n} |\tilde{I}_t|$.
\end{lemma}

\begin{proof}
We introduce an LP to decide which items in $I_B$ will be placed into which containers. 
For some 3D item $i \in I_B$, let $\mathcal{T}_i$ denote the set of types the item can be rounded to.
We introduce variables $x_{i,t}$ that represent, for each \bigy item $i$ and type $t \in \mathcal{T}_i$, the total height of $i$ that is rounded to size $t$.
As $\tilde{I}$ is a rounding that assigns each of the big 3D item slices of height $\frac{\eps v(I^{\infty})}{n}$ to a rounded size, the assignment LP
\begin{align*}
    \sum_{t \in \mathcal{T}_i} x_{i,t} &= h_i &\forall i \in I_B\\
    \sum_{i \in I_B} x_{i,t} &\leq \frac{\eps v(I^{\infty})}{n} |\tilde{I}_t|  &\forall t \in \mathcal{T}\\
    x_{i,t} &\geq 0 &\forall i \in I_B, t \in \mathcal{T}_i
\end{align*}
has a basic solution with at most $|I_B| + |\mathcal{T}|$ non-zero components.
As the number of variables and constraints are $O_{\eps}(1)$, such a solution can be computed in polynomial time.
As each item needs at least one non-zero component at most $|\mathcal{T}| = O_{\eps}(1)$ items are assigned fractionally. 
\end{proof}

For each of the rounded base types $t \in \mathcal{T}$, let $B_t \subseteq I^{\infty}$ denote the set of 3D items that are assigned to have a rounded base $t \in \mathcal{T}$ by the basic solution of the assignment LP.  
We place each of the $|\mathcal{T}|$ fractionally placed \bigy items into individual bins.
In the next step, we assign the items in $B_t$ to containers $K_t$.

\begin{lemma}
\label{lem:placing-big-items-second-step}
By extending the height of one container in $K_t$ by $O(\log(1/\eps))$ and all other containers by $\eps$, it is possible to place all the items $B_t$ into the containers $K_t$ while maintaining the tall-non-sliced property.
\end{lemma}
\begin{proof}
Observe that $h(K_t) \geq h(B_t)$, as the given $(\eps/\mu)$-2D-container-packing is feasible and hence places all items in $I_t$ resulting in $h(B_t) \leq \frac{\eps v(I^{\infty})}{n} |\tilde{I}_t| \leq h(K_t)$.
Let $H_{k}:=\sum_{i=1}^{k}1/i$ denote the $k$-th harmonic number.
Note that $H_{1/\eps} \leq \ln(1/\eps)+1$.
We will build one stack of all items in $B_t$ that fulfills the tall-non-sliced property

The algorithm sorts the items $B_t$ by height and stacks them on top of each other in order of nonincreasing heights. 
Remember that the heights of the tall items in $B_t$ are rounded harmonically, i.e., all items have a height of the form $1/q$ for some $q \in \mathbb{N}$.
For each $q \in \{3,\dots,1/\eps\}$, let $i_q$ be the first item in the stack (from bottom to top) that has a height of $1/q$.
Iterating $q$ from $3$ to $1/\eps$, the algorithm shifts up $i_q$ and all the following items by at most $1/q$ such that the bottom of $i_q$ is aligned to a height that is an integer multiple of $1/q$.
By this step, the height of the stack is increased by at most $H_{1/\eps}-\frac{3}{2}$.
Further, it fulfills the tall-non-sliced property, as items with heights $1/q > \eps$ never overlap integer heights.

In the next step, the algorithm iterates the containers in $K_t$, and for each container $K \in K_t$, it cuts the remaining stack at height $h(K)$ and places all items in this part into the container.  
As all the containers in $K_t$ have an integer height, no tall item is cut in this process. 
If a small item is cut, the rest of the item is placed on top of the container, effectively extending the height of the container by $\eps$.
If $K \in K_t$ is the last container, all the residual items are placed inside it. As the stack has a height of at most $h(K_t) + H_{1/\eps}-\frac{3}{2}$, the height of this last container is increased by at most $O(\log(1/\eps))$.
\end{proof}

Next, we show that the \bigy items can be placed into the containers while maintaining the tall-non-sliced property.

\begin{restatable}{lemma}{bigThreeDContainer}
\label{lem:packing-bins-into-containers}
By extending the height of all containers for \bigy items by $O(\log(1/\eps))$, it is possible to place them into their containers and $O(|\mathcal{T}|)$ additional ones, while maintaining the tall-non-sliced property.
\end{restatable}

\begin{proof}
We place each of the $|\mathcal{T}|$ fractionally placed \bigy items from the proof of \cref{lem:frist-setp-big-items} into individual bins.
In the next step, we assign the items $B_t$ to containers $K_t$ as done by \cref{lem:placing-big-items-second-step}.
As a consequence, all tall items can be placed into their containers whose heights have been extended by $O(\log(1(\eps))$ and $O(|\mathcal{T}|)$ additional ones.
\end{proof}

\noindent \textbf{Placing vertical and horizontal items.}
Note that for horizontal and vertical items, we use the same procedure, except that we rotate the strip and the items by 90 degrees along the $z$-axis. 

For each $d\in \mathcal{D}$, let $\mathcal{V}_d$ denote the set of 3D items whose 2D counterparts have been rounded to be vertical items with rounded depth $d$ and let $K_d$ be the set of all containers for vertical items with rounded depth $d$. 
Since the depth of each container in $K_d$ equals the vertical item depth $d$, we simplify the placement by considering only the $xz$-plane, i.e., the front face of the containers and the items.
Let $\mathcal{V}_{d,xz}$ denote the set of all front faces of items in $\mathcal{V}_d$ and let $K_{d,xz}$ denote the set of front faces of all containers $K_d$ for these vertical items.
\cref{alg:placing-vertical-items-into-containers} specifies how to place almost all 2D items $\mathcal{V}_{d,xz}$ into the 2D containers $K_{d,xz}$.
Note that the y-coordinate of the 3D item corresponds to the y-coordinate of the 3D container.

The idea of the algorithm is to iteratively choose any empty container-face $K \in K_{d,xz}$ and pick a set of 2D items $V_d'$ to be placed that have at least the area of $K$.
To make sure that all items $V_d'$ can be placed into the target area, the container-face $K$ is extended on both sides.
This extension is chosen large enough such that all items $V_d'$ fit, when using the NFDH algorithm to place them.
We use the notation $\texttt{NFDH}(I, a \times b)$ to denote that the set of items $I$ is placed into the rectangular area $a \times b$ with width $a$ and height $b$ using the NFDH algorithm.
The algorithm returns a pair of vectors $(x^N,z^N)$ specifying the x- and z-coordinates of the lower-left corner for each placed item.
In the next step, the items overlapping the width of the container are removed and remembered in a set $V_d''$. 
Later, we will see that the total area of these removed items is not too large.
Remember that the NFDH algorithm places groups of items on shelves.
To ensure the tall-not-sliced property, these shelves are shifted up such that they start at a height that is an integer multiple of the tallest item on the shelf.

\begin{algorithm}
\DontPrintSemicolon
\caption{Algorithm to place the vertical items into containers.}
\label{alg:placing-vertical-items-into-containers}
\textbf{Input:}  $K_{d,xz}, \mathcal{V}_{d,xz}$.\;
\smallskip
Sort the items $\mathcal{V}_{d,xz}$ by height, $V_d'' \leftarrow \emptyset$.\;
\While{$\mathcal{V}_{d,xz} \not =\emptyset$}{
Choose $K \in K_{d,xz}$, $K_{d,xz} \leftarrow K_{d,xz} \setminus \{K\}$, $V_d' \leftarrow 
 \emptyset$.\;
\While{$\area(V_d') < \area(K)$ and $\mathcal{V}_{d,xz} \not = \emptyset $}{
$i \leftarrow \arg \max_{i \in \mathcal{V}_{d,xz}} h_i$, $V_d' \leftarrow V_d'\cup \{i\}$, $\mathcal{V}_{d,xz} \leftarrow \mathcal{V}_{d,xz} \setminus \{i\}$.\;
}
$(x^N,z^N) \leftarrow \texttt{NFDH}(V_d', (w(K) +2\mu) \times (h(K)+2))$.\;
$V_d'' \leftarrow V_d'' \cup \{i \in V_d' \mid w_i+x^N_i > w(K)\}$.\;
Sequentially shift up all NFDH shelves such that every shelf whose tallest item has a height of $1/q > \eps$ is aligned at an integer multiple of $1/q$. Extend the height of the container as needed.
}
\textbf{return} the set of packings into the different areas and $V''_d$.\;
\end{algorithm} 

\begin{lemma}
    \Cref{alg:placing-vertical-items-into-containers} places all vertical items $\mathcal{V}_{d,xz}$ except for $V_d''$ into the containers $K_{d,xz}$ whose heigths have been extended by $H_{1/\eps}+1$.
\end{lemma}
\begin{proof}
First, we prove that it is possible to place all items from $V_d'$ into the area $(w(K) +2\mu) \times (h(K)+2)$. 
As the width of the items in $\mathcal{V}_{d,xz}$ is bounded by $w_{\max} \leq \mu$, and their height is bounded by $h_{\max}\leq 1$, it holds that $\area(V_d') \leq \area(K) +\mu = w(K)h(K)+\mu$. 
By \cref{lem:NFDH}, NFDH places any set of items $V_d'$ into an area $w \times h$ if 
$(w-w_{\max}) \cdot (h -h_{\max}) \geq \area(V_d')$.
This condition is fulfilled as $(w(K) +2\mu -w_{\max}) \cdot (h(K)+2 -h_{\max}) \geq (w(K) +\mu) \cdot (h(K)+1) \geq w(K)h(K)+\mu$.
Therefore, all the items from $V_d'$ can be placed into the area $(w(K) +2\mu) \times (h(K)+2)$.

Next, we show that the algorithm does not run out of containers $K$ before all items in $\mathcal{V}_{d,xz}$ are packed. 
Let $W_d$ be the total width of 2D items with rounded depth $d$ in $\tilde{I}$.
Each of these items corresponds to an item slice with height $\frac{\eps v(I^{\infty})}{n}$. 
Hence, the total area of the items in $\mathcal{V}_{d,xz}$ is bounded by $W_d \cdot \frac{\eps v(I^{\infty})}{n}$.
On the other hand, as we are given a feasible $(\eps/\mu)$-2D-container-packing, we know that the total width of containers for vertical items with rounded depth $d$ is at least $W_d$.
Since each configuration gets assigned a height of at least $\frac{\eps v(I^{\infty})}{n}$, it holds that $\area(K_{d,xz}) \geq \area(\mathcal{V}_{d,xz})$.
In each step, the algorithm selects a set of items $V'_d$ with $\area(V'_d) > \area(K)$, ensuring that containers do not run out before the last item is placed.

Finally, we need to prove that the height $H_{1/\eps}-1$ is a large enough extension to the height of the boxes to ensure that each shelf whose tallest item has a height of $1/q > \eps$ is aligned to a height that is a multiple of $1/q$. Note that as the items are placed in order of their height, there will not be a gap between shelves containing items of height $1$ and items of height $1/2$. Therefore, the total height of gaps created between shelves is bounded by $\sum_{i = 3}^{1/\eps}1/i = H_{1/\eps}-\frac{3}{2} \in O(\log(1/\eps))$.
\end{proof}

Next, we show that iteratively using \cref{alg:placing-vertical-items-into-containers} for all rounded sizes $d \in \mathcal{D}$ gives a packing of all the vertical items into the container and a few additional bins.

\begin{restatable}{lemma}{horizontalThreeDContainer}
Enlarging the height of the 3D vertical containers by $O(\log(1/\eps))$ allows all vertical items to be placed into their containers and at most $O(\eps)(\frac{\eps v(I^{\infty})}{n} B+k)$ additional bins while maintaining the tall-not-sliced property. 
\end{restatable}

\begin{proof}
Note that any solution given by \cref{alg:placing-vertical-items-into-containers} already fulfills the tall-not-sliced property, as all shelves whose tallest item has a height of $1/q > \eps$ is aligned at an integer multiple of $1/q$. Therefore, any item on such a shelf cannot cross any integer height.

It remains to place the items from $V''= \bigcup_{d \in \mathcal{D}}V_d''$.
For any $d \in \mathcal{D}$ and $K \in K_d$, only items that are placed right of $w(K)$ are added to $V_d''$. As each placed item has a width of at most $\mu$, a total width of at most $3\mu$ is removed at each height.
Regarding the height, we added a height of $2$ to each container.  
So the total height of the 3D configurations is bounded by $(\frac{\eps v(I^{\infty})}{n} B+3k)$.
Further, as we consider an $(\eps/\mu)$-2D-container-packing, the total depth of containers for vertical items per configuration is bounded by $O(\eps/\mu)$.
Therefore, the items in $V''$ have a total volume of at most $(O(\eps/\mu) \cdot 3\mu) \cdot (\frac{\eps v(I^{\infty})}{n} B+3k) = O(\eps)(\frac{\eps v(I^{\infty})}{n} B+k)$.
Hence, by \cref{thm:volume-packing}, these items can be placed into $O(\eps)(\frac{\eps v(I^{\infty})}{n} B+k)$ extra bins.
\end{proof}

\noindent \textbf{Placing \tin items.}
The \tin items are placed by iteratively selecting a subset of items and packing them into the containers using the NFDH algorithm.
We show that almost all \tin items can be packed into their designated 3D containers, except for a small-volume subset that can be accommodated in at most $O(\delta)\opt_{\twobp}+O(1)$ additional bins. 

Let $\mathcal{S}$ denote the set of 3D items whose base is classified as \tin by the \twobp algorithm.
Further, let $K_S$ denote the set of containers for \tin items.
As the small items are not rounded to a constant number of sizes in any dimension, we will take a different approach than for the horizontal and vertical items.
We sort the small items by height and iteratively pick a container $K \in K_S$ to fill the items inside.
For this container $K$, we iteratively choose a set of items $\mathcal{S}_{K,\ell}$, for some $\ell \in \mathbb{N}$, by greedily taking the tallest small items until the top faces of the items have a larger area than the area of the top face of the container.
The top-faces $\mathcal{S}_{K,\ell}^{\mathrm{flat}}$ of the items $\mathcal{S}_{K,\ell}$ are then placed into the area $(w(K) +2\mu) \times (d(K)+2\mu))$ using the NFDH algorithm.
In the next step, the items overlapping the container borders are removed and added to the set $\mathcal{S}''$.
The items in $\mathcal{S}_{K,\ell}$ are then actually placed into the container by assigning them their relative $xy$-position in the box and all of them a height such that their top faces align with the same height $h$.
We call the placement of such a set $\mathcal{S}_{K,\ell}$ a layer and they are numbered starting at $1$.
After the last layer intersects the height $h(K)+1$, we close the container. But before we consider the next container, the algorithm ensures the tall-not-sliced property by shifting the layers upwards, such that the z-axis of the layer is aligned with an integer multiple of the tallest item from the layer.

\begin{algorithm}
\DontPrintSemicolon
\caption{Algorithm to place the \tin items into containers.}
\label{alg:placing-tiny-items-into-containers}
\textbf{Input:}  $\mathcal{S}$, $K_S$.\;
\smallskip
Sort the items $\mathcal{S}$ by height, $\mathcal{S}'' \leftarrow \emptyset$.\;
\While{$\mathcal{S} \not =\emptyset$}{
Pick $K \in K_S$, $K_S \leftarrow K_S \setminus \{K\}$, $h \leftarrow 0$, $\ell \leftarrow 0$.\;
\While{$h \leq h(K)+1$ and $\mathcal{S} \not =\emptyset$}{
$\mathcal{S}_{K,\ell} \leftarrow \emptyset$, $\mathcal{S}_{K,\ell}^{\mathrm{flat}} \leftarrow \emptyset$, $\mathcal{S}_{K,\ell}^{\area} \leftarrow 0$, $h \leftarrow h+\max\{h_i | i \in \mathcal{S}\}$.\;
\While{$\mathcal{S}_{K,\ell}^{\area} < w(K) \cdot d(K)$ and $\mathcal{S} \not = \emptyset $}{
$i = \arg \max_{i' \in \mathcal{S}} h_{i'}$,  $\mathcal{S}_{K,\ell} \leftarrow \mathcal{S}_{K,\ell} \cup \{i\}$,  $\mathcal{S}\leftarrow \mathcal{S} \setminus \{i\}$.\;
$\mathcal{S}_{K,\ell}^{\mathrm{flat}}\leftarrow \mathcal{S}_{K,\ell}^{\mathrm{flat}} \cup \{(w_i,d_i)\}$,  $\mathcal{S}_{K,\ell}^{\area} \leftarrow\mathcal{S}_{K,\ell}^{\area} + w_id_i$.\;
}
$(\mathbf{x}^N,\mathbf{y}^N) \leftarrow \texttt{NFDH}(\mathcal{S}_{K,\ell}^{\mathrm{flat}}, (w(K) +2\mu) \times (d(K)+2\mu))$.\;
$\mathcal{S}'' \leftarrow\mathcal{S}'' \cup \{i \in \mathcal{S}_{K,\ell} \mid (x^N_i + w_i > w(K)) \vee (y_i^N + d_i > d(K))\}$.\;
$\forall i \in \mathcal{S}_{K,\ell} \setminus \mathcal{S}'': (x_i,y_i,z_i) \leftarrow (x_K+x^N_i,y_K+y^N_i,h-h_i)$.
}
$h_{s} \leftarrow 0$.\;
\For(\tcp*[f]{ensure tall-non-sliced property}){$\ell' = 1$ \KwTo $\ell$}{
$i =\arg \max_{i' \in \mathcal{S}_{K,\ell'}} h_{i'}$.\;
\If{$h_i > \eps$}{
$q \leftarrow 1/h_i$, $q' \leftarrow \lceil q\cdot (h_{i}+h_{s}+z_i)\rceil$, $h_{s} \leftarrow h_{s} + (q'/q -(h_{i}+h_{s}+z_i))$.\;
$\forall i' \in \mathcal{S}_{K,\ell} \setminus \mathcal{S}'': (x_{i'},y_{i'},z_{i'}) \leftarrow (x_{i'},y_{i'},q'/q-h_{i'})$.
}
}
}
\textbf{return} $(\mathbf{x},\mathbf{y},\mathbf{z}) \in ([0,1)\times[0,1)\times\mathbb{R}_+)^{|\mathcal{S}|}$, $\mathcal{S}''$.\;
\end{algorithm} 

\begin{lemma}
\label{lem:alg-to-place-tiny}
    Given $\mathcal{S}$ and $K_S$, \cref{alg:placing-tiny-items-into-containers}  places all items in $\mathcal{S}$ into the containers $K_S$ whose heights have been extended by $H_{1/\eps}+1$ except for $\mathcal{S}''$ while fulfilling the tall-not-sliced property.
\end{lemma}

\begin{proof}

First, we show that all the upper faces of the items in $\mathcal{S}'$ can be placed into the area $(w(K) +\mu) \times (d(K)+\mu)$. 
By \cref{lem:NFDH}, it is possible to place items with a total area of at least $(w(K) +2\mu-w_{\max}) \times (d(K)+2\mu-d_{\max})$ into the area $(w(K) +2\mu) \times (d(K)+2\mu)$.
As \tin items have a width and a depth that is bounded by $\mu$, NFDH places items with a total volume of at most $w(K)d(K) + \mu(w(k)+d(K))+\mu^2$. 
As each \tin item has an upper face with an area of at most $\mu^2$, it holds that $\mathcal{S}'_{\area} \leq w(K)d(K) +\mu^2$ when placing the items into the rectangular area 
$(w(K) +2\mu) \times (d(K)+2\mu)$.

Next, we show that $\mathcal{S} = \emptyset$ before we try to choose a container $K$ from $ \emptyset$.
As we are given a $(\eps/\mu)$-2D-container packing, the total area of \tin items $S_{\area}$ in the \twobp instance is at most as large as the total area of the containers for small items $K_{S,\area}$. 
As a consequence, $v(K_S) \geq \frac{\eps v(I^{\infty})}{n} K_{S,\area} \geq v(\mathcal{S})$, i.e., the total volume of containers for \tin items is at least as large as the total volume of \tin items.
As for each $\mathcal{S}_{K,\ell'}$, in container $K$, a total area of $w(K)d(K)$ is chosen, it remains to calculate the loss due to the fact that item in $\mathcal{S}_{K,\ell'}$ might have different heights.
Let $h_{K,\ell',\max}$ denote the maximum height in $\mathcal{S}_{K,\ell'}$ and $h_{K,\ell',\min}$ the corresponding minimum height.
Then, due to the property of NFDH, the total volume we lose in $K$ is bounded by $\sum_{\ell' = 1}^L (h_{K,\ell',\max} - h_{K,\ell',\min}) w(K)d(K) \leq \sum_{\ell' = 1}^L (h_{K,\ell',\max} - h_{K,\ell'+1,\max}) w(K)d(K)$, where we define $h_{K,\ell+1,\max}= 0$.
Note that $\sum_{\ell' = 1}^L (h_{\ell',\max} - h_{\ell'+1,\max}) w(K)d(K) = h_{K,1,\max} w(K)d(K) \leq  w(K)d(K)$.
Note that for each container $K$ that is closed before $\mathcal{S} = \emptyset$ in the algorithm, it holds that $h(K) +1 < \sum_{\ell = 1}^L h_{\ell,\max} \leq h(K) +2$. 
As a consequence, the algorithm chooses items with a total volume of at least $(h(K) +1)w(K)d(K) - w(K)d(K) = h(K)w(K)d(K)= v(K)$ to be placed into the container $K$.
As $v(K_S) \geq v(\mathcal{S})$, all items can be placed before the algorithm runs out of containers. 
Note that the height of each container needs to be extended by at most $2 \in O(\log(1/\eps))$ for this process.

Finally, we need to prove that the height $O(\log(1/\eps))$ is a large enough extension to the height of the boxes to ensure that each layer of items $\mathcal{S}_{K,\ell'}$, whose tallest item has a height of $1/q > \eps$, is aligned to a height that is a multiple of $1/q$. Note that as the items are placed in order of their heights, there will not be a gap between layers containing items of height $1$ and items of height $1/2$. Therefore, the total height of gaps created between layers is bounded by $h_s \leq \sum_{q = 3}^{1/\eps}1/q = H_{1/\eps}-\frac{3}{2} \in O(\log(1/\eps))$.
Hence, the height of all containers needs to be extended by at most $O(\log(1/\eps))$ to accommodate all layers.
\end{proof}

It remains to place the items from $\mathcal{S}''$. The next lemma shows that all the \tin items can be placed into their containers and a few additional bins.

\begin{restatable}{lemma}{smallThreeDContainer}
    By enlarging the 3D tiny containers by $O(\log (1/\eps))$ along the height, it is possible to compute a packing of a subset of the tiny items, ensuring the tall-not-sliced property. The remaining tiny items can be packed into $O(\eps)(\frac{\eps v(I^{\infty})}{n}B+k)$ additional bins.
\end{restatable}
\begin{proof}
By \cref{lem:alg-to-place-tiny}, we know that given $\mathcal{S}$ and $K_S$, \cref{alg:placing-tiny-items-into-containers}  places all items in $\mathcal{S}$ into the containers $K_S$ whose heights have been extended by $H_{1/\eps}+1 = O(\log(1/\eps))$, except for $\mathcal{S}''$,  while fulfilling the tall-not-sliced property.
Therefore, it remains to prove that the items from $\mathcal{S}''$ can be placed into $O(\eps)B + O(\eps)(B+k)$ extra bins.
Note that per container $K \in K_S$, the volume attributed to $\mathcal{S}''$ is bounded by
$3\mu \cdot w(K) \cdot (h(K)+2) + 3\mu \cdot d(K) \cdot (h(K)+2) \leq 6\mu (h(K)+2)$.
As the given packing is a $(\eps/\mu)$-2D-container packing, there are at most $O(\eps/\mu)$ containers for \tin items per configuration.
As a consequence, per configuration $C$, we lose a volume of at most $6\mu (h(C)+2) \cdot O(\eps/\mu) = O(\eps)(h(C)+2)$.
Hence the total volume of $\mathcal{S}''$ is bounded by $O(\eps)(\frac{\eps v(I^{\infty})}{n}B+3k) = O(\eps)(\frac{\eps v(I^{\infty})}{n} B+k)$.
Therefore, the items $\mathcal{S}''$ can be placed in $O(\eps)(B+k)$ additional containers.
\end{proof}

These three lemmas about placing \bigy, vertical, horizontal, and \tin items give us the tools to prove \cref{lem:alg-to-put-all-items-into-2D-container-packing}.

\begin{proof}[Proof of \cref{lem:alg-to-put-all-items-into-2D-container-packing}]
Note that for all item types \bigy, horizontal, vertical, and \tin, the containers have to be increased by a height of at most $O(\log(1/\eps))$.
Hence the total height of the 3D configurations is bounded by $\frac{\eps v(I^{\infty})}{n} B+O(\log(1/\eps))k$.
Further, the large items need at most $|\mathcal{T}|$ extra bins while the horizontal and vertical and small items need at most $O(\eps)(\frac{\eps v(I^{\infty})}{n} B+k)$ extra bins.
We use \cref{thm:volume-packing} to place the medium items into $O(\eps)\opt_{\tbp} +O(1) \leq O(\eps)\frac{\eps v(I^{\infty})}{n}B +O(1)$ additional bins, as they have a volume of at most $O(\eps)\frac{\eps v(I^{\infty})}{n}\opt_{\twobp} \leq O(\eps)\opt_{\tbp}(I)$.

In the next step, we transform the packings into the 3D configurations into packings into 3D bins by slicing each configuration at all integer heights. 
The items sliced by this process all have a height of at most $\eps$, and, hence, we can pack items sliced by $1/\eps$ lines inside a single bin. 
As the total height of the configurations is bounded by $\frac{\eps v(I^{\infty})}{n} B+O(\log(1/\eps))k$,
this splitting introduces at most $(1+O(\eps))(\frac{\eps v(I^{\infty})}{n} B+O(\log(1/\eps))k)$ new bins in total.
Therefore, the total number of bins the algorithm produces is bounded by 
$(1+O(\eps))(\frac{\eps v(I^{\infty})}{n} B+O(\log(1/\eps))k) + |\mathcal{T}| + O(\eps)(\frac{\eps v(I^{\infty})}{n} B+k) = (1+O(\eps))\frac{\eps v(I^{\infty})}{n} B + O(\log(1/\eps)k +|\mathcal{T}|)$.
\end{proof}

By \cref{thm:finding-2D-container-packing}, we can find $\tilde{I}$ and a $(1/\delta^3)$-2D-container-packing of $\tilde{I}$ into $B$ bins with at most $O_{\eps}(1)$ different bin-configurations that fulfills $B \leq (\frac{3}{2}+O(\eps))\opt_{\twobp}(I) +O_{\eps}(1)$.
As a consequence, since $\mu \leq \eps \delta^3$, by \cref{lem:alg-to-put-all-items-into-2D-container-packing}, we can find a packing of $I^\infty$ into at most 
$
\left(\frac{3}{2}+O(\eps)\right)\frac{\eps v(I^{\infty})}{n}\cdot \opt_{\twobp}(I_{\twobp}) + O_{\eps}(1)
$
3D bins, because $\frac{\eps v(I^{\infty})}{n} \leq \eps$ and $O(\log(1/\eps)k +|\mathcal{T}|) \in O_{\eps}(1)$.
Since by \cref{lem:asymp-harmonic} and \cref{lem:asymp-tsp-to-twobsp}, we have that 
$\frac{\eps v(I^{\infty})}{n}\opt_{\twobp}(I_{\twobp})  
\leq (1+\eps)\opt_{\tsp}(I^{\infty}) 
\leq (1+\eps)T_{\infty}\opt_{\tbp}(I)$,
the solution generated by the algorithm uses at most $(\frac{3}{2}+O(\eps))T_{\infty}\cdot \opt_{\tbp}(I) + O_{\eps}(1)$ bins
establishing \Cref{thm:3dbpasymp}.



\section{Algorithms for 3D-MVBB}
\label{sec:minvolumecontainer}

In this section, we establish \Cref{thm:mvbb}. 
First, we guess the dimensions of the optimal bounding box within factors of $1+\varepsilon$ and then scale the dimensions of each item by the corresponding (guessed) dimensions of the bounding box so that all items now fit inside a $1\times 1 \times 1$ bin. 

\subsection{Absolute approximation} 
We first compute a value of $\mu$ in order to classify the items using the following lemma.

\begin{lemma}
\label{lem:mediumitems}
     There exists a polynomial-time computable $\mu \le \varepsilon$ such that the total volume of the (scaled) items having at least one of the dimensions in the range $(\mu^6,\mu]$ is at most $\varepsilon$.
\end{lemma}
\begin{proof}
    Similar to the proof of \Cref{lem:medium}, we define $\mu_0 = \eps$, and for $j\in [3/\eps]$, define $\mu_j = \mu_{j-1}^6$. Let $I_j$ be the items with one of the dimensions lying in the range $(\mu_j,\mu_{j-1}]$. Since $v(I)\le 1$, and each item belongs to at most three of the sets $\{I_j\}_{j\in [3/\eps]}$, we have $\sum_{j\in [3/\eps]} v(I_j) \le 3$, and therefore there exists a $j^*$ for which $v(I_{j^*}) \le \eps$. We set $\mu = \mu_{j^*-1}$ and are done.
\end{proof}

We now classify the items as follows -- let $L$ be the items whose height, width and depth all exceed $\mu$, which we will refer to as \emph{large} items. Let $I_h$ be the items with height at most $\mu^6$, $I_w$ be the remaining items with width at most $\mu^6$ and $I_d$ be the remaining items having depth at most $\mu^6$. Finally, let $I^{\text{rem}}$ be the items having at least one dimension in the range $(\mu^6,\mu]$, whose total volume is bounded by $\varepsilon$ by \Cref{lem:mediumitems}. These items are further subdivided as follows: let $I^{\text{rem}}_h \subseteq I^{\text{rem}}$ be the items with height at most $\mu$, $I^{\text{rem}}_w \subseteq I^{\text{rem}} \setminus I^{\text{rem}}_h$ be the items with width at most $\mu$, and $I^{\text{rem}}_d = I^{\text{rem}}\setminus (I^{\text{rem}}_h \cup I^{\text{rem}}_w)$ be the remaining items with depth at most $\mu$.

\begin{restatable}{lemma}{packmedium}
\label{lem:packmedium}
    The items of $I^{\text{rem}}_h$ (resp. $I^{\text{rem}}_w,I^{\text{rem}}_d$) can be packed inside a strip with $1\times 1$ base and height (resp. width, depth) at most $12\varepsilon$.
\end{restatable}
\begin{proof}
    Since $v(I^{\text{rem}}_h) \le \varepsilon$ and the height of each item in $I^{\text{rem}}_h$ is bounded by $\mu$, they can be packed within a height of $4\eps + 8\mu \le 12\varepsilon$ by \Cref{thm:licheng}. Analogous claims hold for $I^{\text{rem}}_w$ and $I^{\text{rem}}_d$.
\end{proof}

We now consider the packing of the items in $L\cup I_h$ and obtain the following result.
\begin{restatable}{proposition}{packLandIh}
\label{pro:packLandIh}
    In polynomial time, we can compute a packing of $L\cup I_h$ inside a cubic bin of side length $1+O(\varepsilon)$.
\end{restatable}


\begin{proof}
    We classify the items of $I_h$ based on their width and depth into big, horizontal, vertical and small items as follows -- an item $i$ is \emph{big} if both $w_i, d_i > \mu$, \emph{horizontal} if $w_i > \mu$ and $d_i \le \mu^6$, \emph{vertical} if $d_i > \mu$ and $w_i \le \mu^6$ and \emph{tiny} if both $w_i, d_i \le \mu^6$. Using a result of Bansal et al.~\cite{bansal2006bin}, who gave an APTAS for \twobp~when the sides of each bin can be augmented by $1+O(\varepsilon)$, we can discretize the $x$- and $y$-coordinates of all items of $L$ and the non-tiny items of $I_h$.

\begin{lemma}[\cite{bansal2006bin}]
\label{lem:bansaldiscretize}
    By enlarging the width and depth of the bin by $\mu$, we can ensure that the width and the $x$-coordinate of any item with $w_i > \mu$ is rounded up to an integer multiple of $\mu^2/2$, and the depth and the $y$-coordinate of any item with $d_i > \mu$ is rounded up to an integer multiple of $\mu^2/2$.
\end{lemma}

In a similar way, by extending the height of the bin by $\mu$, we can discretize the heights of the large items. Also, we can discretize the $z$-coordinates of all the items.

\begin{lemma}
\label{lem:discretizeposition}
    Assume $\mu n$ is an integer so that $\mu^2$ is an integer multiple of $\mu/n$. By enlarging the height of the bin by $2\mu$, we can assume the following.
    \begin{itemize}
        \item The heights of all items are integer multiples of $\mu/n$, and they are positioned at $z$-coordinates that are integer multiples of $\mu/n$.
        \item The heights of the large items are rounded up to integer multiples of $\mu^2/2$. Further, they are placed at $z$-coordinates that are integer multiples of $\mu^2/2$.
    \end{itemize}
\end{lemma}

Since there are only at most $1/\mu^3$ large items and $O(1/\mu^6)$ possible positions for them inside the bin by Lemmas \ref{lem:bansaldiscretize} and \ref{lem:discretizeposition}, we can guess all of them in $O_{\mu}(1)$ time. Assuming we have guessed all of them correctly, we now have a bin with at most $1/\mu^3$ holes, and our goal is to pack the items of $I_h$ in the space outside the holes. For this, we adopt the same approach as described in \Cref{sec:asymptotic}. We slice the items of $I_h$ at heights of $\mu/n$, thus obtaining an instance of \twobp, where the bins now have holes. By Lemmas \ref{lem:bansaldiscretize} and \ref{lem:discretizeposition}, the positions of the items inside the (enlarged) bins are already discretized, and thus the number of possible 3D configurations is only $O(1/\mu^4)$.

It remains to pack the items of $I_h$ into their respective 3D containers. This is done similar to \cite{3d-strip-packing}. First the height of each configuration is extended by the maximum height of an item in $I_h$, which is bounded by $\mu^6$. Since the total number of configurations is bounded by $O(1/\mu^4)$, the increase in the height of the packing is at most $O(\mu^2)$. Using a result of Lenstra et al.~\cite{lenstra1990approximation}, all the big items can be packed inside the extended big containers (extended by $\mu^6$ along the height) by stacking them one above the other. The horizontal items of a particular rounded width are packed into horizontal containers of the same width using NFDH. For this, the depth of the container is extended by $\mu^6$, and the horizontal items are packed into the extended containers using NFDH. Analogously, the vertical items are packed into extended vertical containers. For packing tiny items into tiny containers, again the width and depth of the container are extended by $\mu^6$, and the items are packed using 3D-NFDH (\Cref{alg:placing-tiny-items-into-containers}). 

Finally, it remains to remove the extensions of the containers and pack the items intersecting the extensions at the top of all the configurations. This is done similar to \cite{3d-strip-packing}, which shows that the extensions can be packed within a height of $O(\mu^2)$. Thus the overall height of the packing of $L\cup I_h$ is at most $1+2\mu + O(\mu^2) \le 1+O(\varepsilon)$.
\end{proof}

As a special case, \Cref{pro:packLandIh} applied separately to $I_w$ and $I_d$ also yields packings into cubic bins of side length $1+O(\varepsilon)$. Together with \Cref{lem:packmedium}, we obtain a packing of all input items into a bounding box of volume $3+O(\varepsilon)$, establishing the absolute approximation guarantee of \Cref{thm:mvbb}.


\subsection{APTAS}
The technique described in the previous subsection (\Cref{lem:discretizeposition} together with \cite{3d-strip-packing}) gives an APTAS for \tsp~with resource augmentation.

\begin{lemma}[\cite{bansal2006bin,3d-strip-packing}]
    For any $\varepsilon>0$, it is possible to compute a packing into a strip whose width and depth are augmented by a factor of $1+O(\varepsilon)$ and whose height is at most $(1+\varepsilon)\opt_{\tsp}+\varepsilon+O_{\varepsilon}(1)h_{\max}$.
\end{lemma}

We run the above algorithm with the base of the strip aligned along the $xy$-, $yz$- and $zx$-planes, respectively, and return the bounding box with the minimum volume. Let $\delta$ be a small constant depending on $\varepsilon$. Note that clearly, the height, width, and depth of the optimal bounding box must be at least $h_{\max}, w_{\max}$ and $d_{\max}$, respectively. Therefore, the volume of the optimal box is lower bounded by $h_{\text{max}}w_{\text{max}}d_{\text{max}}$. 
In the asymptotic case, we can assume $\opt >> h_{\text{max}} w_{\text{max}} d_{\text{max}}$.
Hence, in order to establish an APTAS, we need that the volume of the box returned by our algorithm is at most $(1+O(\eps))\opt_{\tmvc}+O_{\eps}(1)h_{\text{max}}w_{\text{max}}d_{\text{max}}$. We divide the analysis into two cases depending on the values of $h_{\text{max}}, w_{\text{max}}, d_{\text{max}}$.

\textbf{Case 1: $h_{\text{max}}, w_{\text{max}}, d_{\text{max}}$ all exceed $\delta$ -} Consider the Strip Packing along the $z$-axis. Since $w_{\text{max}}, d_{\text{max}} > \delta$, the height of the packing is at most $1+2\varepsilon+O_{\varepsilon}(1)\cdot (1/\delta^2)h_{\text{max}}w_{\text{max}}d_{\text{max}}$. Hence, the volume of the bounding box obtained is also at most $1+O(\varepsilon)+O_{\varepsilon}(1)h_{\text{max}}w_{\text{max}}d_{\text{max}}$, and we have an APTAS for this case.

\textbf{Case 2: At least one out of $h_{\text{max}}, w_{\text{max}}, d_{\text{max}}$ does not exceed $\delta$ -} W.l.o.g.~assume that $h_{\text{max}}\le \delta$. Then the packing along the $z$-axis will have a height bounded by $1+2\varepsilon+O_{\varepsilon}(1)\cdot \delta \le 1+3\varepsilon$, for sufficiently small $\delta$. Therefore, the bounding box also has a volume of at most $1+O(\varepsilon)$, and we, in fact, obtain a PTAS in this case.

Overall, we have the following theorem.

\begin{theorem}
    For any $\eps >0$, there is a polynomial-time algorithm that returns a packing into a bounding box of volume $(1+\eps)\opt_{\tmvc}+O_{\eps}(1)h_{\max}w_{\max}d_{\max}$.
\end{theorem}



\section{3D-BP with rotation}
\label{sec:rotation}


In this section, we show the following result.

\begin{theorem}
    There exists a polynomial-time 5-approximation for \tbp~with rotations.
\end{theorem}
\begin{proof}
    The work of Sharma \cite{Sharma21} gave a $(T_{\infty}^2+\eps)$-asymptotic approximation for \tbp~with rotations, building on the algorithm of Caprara \cite{caprara2008packing}. Let $K$ be a large constant so that the algorithm of Sharma gives an absolute approximation ratio of 5 when $\opt_{\tbp} > K$. From now on, we assume $\opt_{\tbp} = k \le K$ and that we have correctly guessed the value of $k$. Our goal is to obtain a packing of all items using at most $5k$ bins. Define $\mu = \frac{1}{12^4K}$ and let $L$ be the items having all dimensions exceeding $\mu$, which we will refer to as the \emph{large} items from now on.

    We greedily create at most $4k$ groups of items from $I\setminus L$, where each group has a volume in $[1/4-3\mu, 1/4-2\mu]$ except for the last one, which can have a smaller volume if it contains all residual items from $I\setminus L$.
    Finding such groups is possible as one dimension of each item in $I\setminus L$ is bounded by $\mu$, and hence the volume of each item in $I\setminus L$ is bounded by $\mu$.
    Each such group can be packed into a single bin using \Cref{thm:licheng} since $4(1/4-2\mu)+8\mu = 1$. 
    
    If all items of $I\setminus L$ are already packed into the first $4k$ bins, we pack the items of $L$ into an additional $k$ bins by brute-force enumeration. Thus we obtain a packing of $I$ into $5k$ bins, and are done. 
    
    Otherwise, we have packed a volume of at least $(1-12\mu)k$ into the first $4k$ bins. 
    Let $T$ denote the set of unpacked items of $I\setminus L$. Note that since $\opt_{\tbp} = k$, we have $v(L\cup T) \le 12\mu k$.

    \begin{claim}
        One of the dimensions of each item of $L$ does not exceed $1/12$.
    \end{claim}
    \begin{proof}
        Since $v(L) \le 12\mu k \le 1/12^3$, every large item must have a side of length at most $1/12$.
    \end{proof}

    We orient each large item so that its height is at most $1/12$. Using \Cref{thm:licheng}, the large items can be packed inside a strip with $1\times 1$ base and height $4\cdot 12\mu k + 8/12 < 3/4$. Finally, the items of $T$ are also packed in a strip of height at most $4\cdot 12\mu k + 8\mu < 1/4$ using \Cref{thm:licheng}. Thus, the items in $L\cup T$ can all be packed inside a single bin. Overall, we obtain a packing of $I$ into $4k+1 \le 5k$ bins and are done.
\end{proof}

\section{Conclusion}
\label{sec:conc}
We obtained an improved absolute and asymptotic approximation algorithms for \tbp and \tmvc.
Our framework is quite general and should extend to other cases.
E.g., for the case with rotations, we expect that our techniques should be easily extendable to provide similar asymptotic guarantees. 
We also expect that the asymptotic approximation algorithm for $\tbp$  should extend to $d$-dimensions ($d>3$) and provide a $3T_{\infty}^{d-2}/2$-approximation. 
The existence of a PTAS (or hardness) for \tmvc is still open. 
It is also interesting to obtain improved guarantees in pseudopolynomial-time (when the input numeric data is polynomially bounded in $n$).

\bibliography{ref}

@misc{optil,
  title = {{OPTIL}.io},
  howpublished = {\url{https://www.optil.io/optilion/problem/3017}},
}

@misc{vmac,
  title = {{ICRA} Virtual Manufacturing Automation Competition},
  howpublished = {\url{https://www.icra2013.org/index0c2d.html?page_id=231}},
}

@article{meir1968packing,
  title={On packing of squares and cubes},
  author={Meir, Aram and Moser, Leo},
  journal={Journal of Combinatorial Theory},
  volume={5},
  number={2},
  pages={126--134},
  year={1968},
  publisher={Elsevier}
}

@article{coffman1980performance,
  title={Performance bounds for level-oriented two-dimensional packing algorithms},
  author={Coffman, Jr, Edward G and Garey, Michael R and Johnson, David S and Tarjan, Robert Endre},
  journal={SIAM Journal on Computing},
  volume={9},
  number={4},
  pages={808--826},
  year={1980},
  publisher={SIAM}
}

@article{lu2015packing,
  title={Packing cubes into a cube is NP-complete in the strong sense},
  author={Lu, Yiping and Chen, Danny Z and Cha, Jianzhong},
  journal={Journal of Combinatorial Optimization},
  volume={29},
  pages={197--215},
  year={2015},
  publisher={Springer}
}

@article{LeeL85,
  author       = {C. C. Lee and
                  D. T. Lee},
  title        = {A Simple On-Line Bin-Packing Algorithm},
  journal      = {Journal of the ACM},
  volume       = {32},
  number       = {3},
  pages        = {562--572},
  year         = {1985}
}

@article{ferreira1999packing,
  title={Packing squares into squares},
  author={Ferreira, Carlos E and Miyazawa, Flavio K and Wakabayashi, Yoshiko},
  journal={Pesquisa Operacional},
  volume={19},
  number={2},
  pages={223--237},
  year={1999}
}

@article{gilmore1965multistage,
  title={Multistage cutting stock problems of two and more dimensions},
  author={Gilmore, Paul C and Gomory, Ralph E},
  journal={Operations Research},
  volume={13},
  number={1},
  pages={94--120},
  year={1965},
  publisher={INFORMS}
}

@inproceedings{Sharma21,
  author       = {Eklavya Sharma},
  title        = {Harmonic Algorithms for Packing d-Dimensional Cuboids into Bins},
  booktitle    = {FSTTCS},
  pages        = {32:1--32:22},
  year         = {2021}
}

@article{Harren09,
  author       = {Rolf Harren},
  title        = {Approximation algorithms for orthogonal packing problems for hypercubes},
  journal      = {Theoretical Computer Science},
  volume       = {410},
  number       = {44},
  pages        = {4504--4532},
  year         = {2009}
}

@article{csirik1993line,
  title={An on-line algorithm for multidimensional bin packing},
  author={Csirik, J{\'a}nos and Van Vliet, Andr{\'e}},
  journal={Operations Research Letters},
  volume={13},
  number={3},
  pages={149--158},
  year={1993},
  publisher={Elsevier}
}

@inproceedings{Jansen0LS22,
  author       = {Klaus Jansen and
                  Arindam Khan and
                  Marvin Lira and
                  K. V. N. Sreenivas},
  title        = {A {PTAS} for Packing Hypercubes into a Knapsack},
  booktitle    = {ICALP},
  pages        = {78:1--78:20},
  year         = {2022}
}

@InProceedings{3d-strip-packing,
author="Jansen, Klaus
and Pr{\"a}del, Lars",
title="A New Asymptotic Approximation Algorithm for 3-Dimensional Strip Packing",
booktitle="SOFSEM",
year="2014",
pages="327--338"
}

@article{li-cheng,
  title={On three-dimensional packing},
  author={Li, Keqin and Cheng, Kam-Hoi},
  journal={SIAM Journal on Computing},
  volume={19},
  number={5},
  pages={847--867},
  year={1990},
  publisher={SIAM}
}

@article{caprara2008packing,
  title={Packing d-dimensional bins in d stages},
  author={Caprara, Alberto},
  journal={Mathematics of Operations Research},
  volume={33},
  number={1},
  pages={203--215},
  year={2008},
  publisher={INFORMS}
}

@article{2dknapsack-lpacking,
  title={Approximating geometric knapsack via {L}-packings},
  author={G{\'a}lvez, Waldo and Grandoni, Fabrizio and Ingala, Salvatore and Heydrich, Sandy and Khan, Arindam and Wiese, Andreas},
  journal={ACM Transactions on Algorithms},
  volume={17},
  number={4},
  pages={1--67},
  year={2021},
  publisher={ACM New York, NY}
}

@article{martello2000three,
  title={The three-dimensional bin packing problem},
  author={Martello, Silvano and Pisinger, David and Vigo, Daniele},
  journal={Operations Research},
  volume={48},
  number={2},
  pages={256--267},
  year={2000},
  publisher={INFORMS}
}

@article{lodi2002heuristic,
  title={Heuristic algorithms for the three-dimensional bin packing problem},
  author={Lodi, Andrea and Martello, Silvano and Vigo, Daniele},
  journal={European Journal of Operational Research},
  volume={141},
  number={2},
  pages={410--420},
  year={2002},
  publisher={Elsevier}
}

@article{faroe2003guided,
  title={Guided local search for the three-dimensional bin-packing problem},
  author={Faroe, Oluf and Pisinger, David and Zachariasen, Martin},
  journal={INFORMS journal on Computing},
  volume={15},
  number={3},
  pages={267--283},
  year={2003},
  publisher={INFORMS}
}

@article{parreno2010hybrid,
  title={A hybrid GRASP/VND algorithm for two-and three-dimensional bin packing},
  author={Parre{\~n}o, Francisco and Alvarez-Vald{\'e}s, Ram{\'o}n and Oliveira, Jos{\'e} Fernando and Tamarit, Jos{\'e} Manuel},
  journal={Annals of Operations Research},
  volume={179},
  pages={203--220},
  year={2010},
  publisher={Springer}
}

@article{crainic2009ts2pack,
  title={TS2PACK: A two-level tabu search for the three-dimensional bin packing problem},
  author={Crainic, Teodor Gabriel and Perboli, Guido and Tadei, Roberto},
  journal={European Journal of Operational Research},
  volume={195},
  number={3},
  pages={744--760},
  year={2009},
  publisher={Elsevier}
}

@article{crainic2008extreme,
  title={Extreme point-based heuristics for three-dimensional bin packing},
  author={Crainic, Teodor Gabriel and Perboli, Guido and Tadei, Roberto},
  journal={INFORMS Journal on Computing},
  volume={20},
  number={3},
  pages={368--384},
  year={2008},
  publisher={INFORMS}
}

@inproceedings{epstein2004optimal,
  title={Optimal online bounded space multidimensional packing},
  author={Epstein, Leah and van Stee, Rob},
  booktitle={SODA},
  pages={214--223},
  year={2004}
}

@article{chlebik2009hardness,
  title={Hardness of approximation for orthogonal rectangle packing and covering problems},
  author={Chleb{\'\i}k, Miroslav and Chleb{\'\i}kov{\'a}, Janka},
  journal={Journal of Discrete Algorithms},
  volume={7},
  number={3},
  pages={291--305},
  year={2009},
  publisher={Elsevier}
}

@inproceedings{bansal2014improved,
  title={Improved approximation algorithm for two-dimensional bin packing},
  author={Bansal, Nikhil and Khan, Arindam},
  booktitle={SODA},
  pages={13--25},
  year={2014}
}

@article{christensen2016multidimensional,
  author    = {Henrik I Christensen and  Arindam Khan and Sebastian Pokutta and Prasad Tetali},
  title     = {Approximation and online algorithms for multidimensional bin packing: A survey},
  journal   = {Computer Science Review},
  volume    = {24},
  pages     = {63--79},
  year      = {2017}
}

@article{3d-knapsack-diedrich,
  title={Approximation algorithms for 3D orthogonal knapsack},
  author={Diedrich, Florian and Harren, Rolf and Jansen, Klaus and Th{\"o}le, Ralf and Thomas, Henning},
  journal={Journal of Computer Science and Technology},
  volume={23},
  number={5},
  pages={749--762},
  year={2008},
  publisher={Springer}
}

@article{alt2016computational,
  title={Computational aspects of packing problems},
  author={Alt, Helmut},
  journal={Bulletin of EATCS},
  volume={1},
  number={118},
  year={2016}
}

@article{harren2013two,
  title={Two for one: tight approximation of 2d bin packing},
  author={Harren, Rolf and Jansen, Klaus and Praedel, Lars and Schwarz, Ulrich M and van Stee, Rob},
  journal={International Journal of Foundations of Computer Science},
  volume={24},
  number={08},
  pages={1299--1327},
  year={2013},
  publisher={World Scientific}
}

@article{bansal2010new,
  title={A new approximation method for set covering problems, with applications to multidimensional bin packing},
  author={Bansal, Nikhil and Caprara, Alberto and Sviridenko, Maxim},
  journal={SIAM Journal on Computing},
  volume={39},
  number={4},
  pages={1256--1278},
  year={2010},
  publisher={SIAM}
}

@article{jansen2016new,
  title={New approximability results for two-dimensional bin packing},
  author={Jansen, Klaus and Pr{\"a}del, Lars},
  journal={Algorithmica},
  volume={74},
  pages={208--269},
  year={2016},
  publisher={Springer}
}

@article{bansal2006bin,
  title={Bin packing in multiple dimensions: inapproximability results and approximation schemes},
  author={Bansal, Nikhil and Correa, Jos{\'e} R and Kenyon, Claire and Sviridenko, Maxim},
  journal={Mathematics of Operations Research},
  volume={31},
  number={1},
  pages={31--49},
  year={2006},
  publisher={INFORMS}
}

@article{li1992heuristic,
  title={Heuristic algorithms for on-line packing in three dimensions},
  author={Li, Keoin and Cheng, Kam Hoi},
  journal={Journal of Algorithms},
  volume={13},
  number={4},
  pages={589--605},
  year={1992},
  publisher={Elsevier}
}

@inproceedings{miyazawa2004packing,
  title={Packing problems with orthogonal rotations},
  author={Miyazawa, Flavio Keidi and Wakabayashi, Yoshiko},
  booktitle={LATIN},
  pages={359--368},
  year={2004}
}

@article{miyazawa1997algorithm,
  title={An algorithm for the three-dimensional packing problem with asymptotic performance analysis},
  author={Miyazawa, Flavio Keidi and Wakabayashi, Yoshiko},
  journal={Algorithmica},
  volume={18},
  number={1},
  pages={122--144},
  year={1997},
  publisher={Springer}
}

@inproceedings{jansen2006asymptotic,
  title={An asymptotic approximation algorithm for 3 d-strip packing},
  author={Jansen, Klaus and Solis-Oba, Roberto},
  booktitle={SODA},
  pages={143--152},
  year={2006}
}

@article{bansal2007harmonic,
author = {Bansal, Nikhil and Han, Xin and Iwama, Kazuo and Sviridenko, Maxim and Zhang, Guochuan},
title = {A Harmonic Algorithm for the 3D Strip Packing Problem},
journal = {SIAM Journal on Computing},
volume = {42},
number = {2},
pages = {579--592},
year = {2013}
}

@article{hifi2010linear,
  title={A linear programming approach for the three-dimensional bin-packing problem},
  author={Hifi, Mhand and Kacem, Imed and N{\`e}gre, St{\'e}phane and Wu, Lei},
  journal={Electronic Notes in Discrete Mathematics},
  volume={36},
  pages={993--1000},
  year={2010},
  publisher={Elsevier}
}

@article{jin2003three,
  title={The three-dimensional bin packing problem and its practical algorithm},
  author={Jin, Zhihong and Ito, Takahiro and Ohno, Katsuhisa},
  journal={JSME International Journal Series C Mechanical Systems, Machine Elements and Manufacturing},
  volume={46},
  number={1},
  pages={60--66},
  year={2003},
  publisher={The Japan Society of Mechanical Engineers}
}

@inproceedings{li2014genetic,
  title={A genetic algorithm for the three-dimensional bin packing problem with heterogeneous bins},
  author={Li, Xueping and Zhao, Zhaoxia and Zhang, Kaike},
  booktitle={IIE Annual Conference},
  pages={2039},
  year={2014}
}

@article{mahvash2018column,
  title={A column generation-based heuristic for the three-dimensional bin packing problem with rotation},
  author={Mahvash, Batoul and Awasthi, Anjali and Chauhan, Satyaveer},
  journal={Journal of the Operational Research Society},
  volume={69},
  number={1},
  pages={78--90},
  year={2018},
  publisher={Taylor \& Francis}
}

@article{mack2012heuristic,
  title={A heuristic for solving large bin packing problems in two and three dimensions},
  author={Mack, Daniel and Bortfeldt, Andreas},
  journal={Central European Journal of Operations Research},
  volume={20},
  pages={337--354},
  year={2012},
  publisher={Springer}
}

@article{alt2018approximating,
  title={Approximating smallest containers for packing three-dimensional convex objects},
  author={Alt, Helmut and Scharf, Nadja},
  journal={International Journal of Computational Geometry \& Applications},
  volume={28},
  number={2},
  pages={111--128},
  year={2018},
  publisher={World Scientific}
}

@article{ali2022line,
  title={On-line three-dimensional packing problems: A review of off-line and on-line solution approaches},
  author={Ali, Sara and Ramos, Ant{\'o}nio Galr{\~a}o and Carravilla, Maria Ant{\'o}nia and Oliveira, Jos{\'e} Fernando},
  journal={Computers \& Industrial Engineering},
  volume={168},
  pages={108--122},
  year={2022},
  publisher={Elsevier}
}

@article{lenstra1990approximation,
  title={Approximation algorithms for scheduling unrelated parallel machines},
  author={Lenstra, Jan Karel and Shmoys, David B and Tardos, {\'E}va},
  journal={Mathematical Programming},
  volume={46},
  pages={259--271},
  year={1990},
  publisher={Springer}
}

@article{fleischer2011tight,
  title={Tight approximation algorithms for maximum separable assignment problems},
  author={Fleischer, Lisa and Goemans, Michel X and Mirrokni, Vahab S and Sviridenko, Maxim},
  journal={Mathematics of Operations Research},
  volume={36},
  number={3},
  pages={416--431},
  year={2011},
  publisher={Informs}
}

@article{VegaL81,
  author       = {Wenceslas Fernandez de la Vega and
                  George S. Lueker},
  title        = {Bin packing can be solved within 1+epsilon in linear time},
  journal      = {Combinatorica},
  volume       = {1},
  number       = {4},
  pages        = {349--355},
  year         = {1981}
}

@article{steinberg1997strip,
  title={A strip-packing algorithm with absolute performance bound 2},
  author={Steinberg, A},
  journal={SIAM Journal on Computing},
  volume={26},
  number={2},
  pages={401--409},
  year={1997},
  publisher={SIAM}
}

@article{buchwald2016upper,
  title={Upper bounds for heuristic approaches to the strip packing problem},
  author={Buchwald, Torsten and Scheithauer, Guntram},
  journal={International Transactions in Operational Research},
  volume={23},
  number={1-2},
  pages={93--119},
  year={2016},
  publisher={Wiley Online Library}
}
\appendix

\section{Structural result for 2D-BP}
\label{subsec:2dbpAppndix}
In this section, we focus on the \twobp algorithm, and we restate all the ingredients to prove the main structure lemma for \twobp from \cite{jansen2016new}.

\twoBPStructure*

The proof works in three steps. First, we prove that a suitable rounding of the items exists, and then we prove that each packing of the rounded items can be rearranged into a $(1/\delta^3)$-2D-container-packing. Finally, we present an algorithm to find such a $(1/\delta^3)$-2D-container-packing.

\subsection{Existence of a suitable rounding for the items in 2D-BP}
To round the items, it will become necessary to slice horizontal items horizontally (along $x$-axis) and vertical items vertically (along $y$-axis). 
Later, when constructing the container packing, it will become necessary to slice the \tin items in any direction.
We denote by $\opt_{\twobp}^{\mathrm{slice}}(I)$ an optimal solution for a \twobp instance $I$, where we allow this kind of slicing of the items. 
In this section, we prove the following lemma, which states that there exists a suitable rounding of the items of the given \twobp instance.

\begin{lemma}
\label{lem:twobp-rounding-main}
Given $\eps > 0$ and  a set of items $I$ and an optimal packing into $\opt_{\twobp}(I)$ bins, we can find a rounded instance $\tilde{I}$
where the \bigy items are rounded into $1/\delta^4$ types $\mathcal{T}$ such that either the width or the depth is a multiple of $\delta^2$, the vertical items have  $O(1/\eps \cdot \log(1/\delta))$ different depths $\mathcal{D}$, and the horizontal items have at most $O(1/\eps \cdot \log(1/\delta))$ different widths $\mathcal{W}$, such that $\opt_{\twobp}^{\mathrm{slice}}(\tilde{I}) \leq ((3/2)+O(\eps))\opt_{\twobp}(I) + O(1)$.
The rounded instance $\tilde{I}$ is one of $n^{O_{\eps}(1)}$ possibilities to round the items.
\end{lemma}

First, note that we can find the right value $\delta$ to classify the items by size in polynomial time. Further, we can assume that $1/\eps$ and $1/\delta$ are integer values.

\begin{lemma}[\cite{jansen2016new}]
\label{lem:delta-twobp}
    There exists a polynomial-time computable $\delta \in [\eps^{O_{\eps}(1)},\eps]$ such that the total area of items with width in $[\delta^4,\delta)$ or depth in $[\delta^4,\delta)$ is bounded by $\eps\cdot \area(I)$ and $1/\delta$ is a multiple of $24$.
\end{lemma}

\paragraph*{Rounding vertical and horizontal items}
Rounding the width of horizontal items and the height of vertical items is classically done by linearly or geometrically grouping the items. 
Indeed, this is also possible when considering \twobp. 
For completeness, we add the proof here.
The rounding will lead to a total area of at most $O(\eps)\opt_{\twobp}$ of vertical and horizontal items that cannot be placed inside the original bins. 
The main work is to argue that these items can be placed into $O(\eps)\opt_{\twobp} +O(1)$ additional bins.

Note that the rounding can be adjusted such that all the slices of a 3D item whose top faces have been classified as vertical or horizontal are rounded to the same depth or height, respectively.

\begin{lemma}
\label{lem:rounding-vert-hor-items-2bp}
Let any packing of the items $I$ into $B$ bins be given.
Assume we are allowed to slice the vertical items vertically and the horizontal items horizontally.
Then we can find a packing into at most $B + O(\eps \opt_{\twobp}) +O(1)$ bins, where the depths of vertical and widths of horizontal items are rounded to $O(\frac{1}{\eps}\log(\frac{1}{\delta}))$ sizes.
\end{lemma}
\begin{proof}
The main idea to round the items is to use geometric grouping, introduced by~\cite{VegaL81}.
We do the rounding exemplarily for horizontal items, but the rounding for vertical items works analogously when flipping depth and width.
The first step is to group the horizontal items by width into groups $G_i := \{i \in I| d_i< \delta, 1/2^i < w_i \leq 1/2^{i-1}] \}$, 
for each $i \in \{1,2,\dots,\lceil \log(1/\delta) \rceil\}$. In the next step, for each group $G_i$, a linear grouping step rounds the width in this group to $2/\eps$ sizes:
All items in group $G_i$ are sorted by width and stacked on top of each other such that the widest item is at the bottom. Let $h(G_i)$ be the total depth of this stack, and let $\sigma(i)$ denote the depth at which the item starts in the stack.
The stack is now partitioned into subgroups $G_{i,j}$ for each $j \in\{1,\dots, \lceil2/\eps\rceil\}$ such that $G_{i,j}$ contains all items with $(j-1)\eps h(G_i)/2 < \sigma(i) +i_h \leq j\eps h(G_i)/2$.
Items overlapping some multiple of $j\eps h(G_i)/2$ are sliced at this line.
In the rounding step, the widths of all the items in a group $G_{i,j}$ are rounded to the width of the widest item in the group, except for the items that have the smallest width inside the group, 
which will keep their width, if it is the same as the widest width in the next group $G_{i,j+1}$. 
Note that when allowing to horizontally slice items, the rounded items from group $G_{i,j}$, for each $j> 1$, can be placed instead of the rounded items from group $G_{i,j-1}$, since the widest original width in group $G_{i,j-1}$ has a width that is at least as large as the rounded width of each item in $G_{i,j}$.

However, the items from the group $G_{i,1}$ cannot be placed in the place of other items from any other group. 
These items will be placed into new bins.
Note that the items from group $G_{i,1}$ have a width of at most $\frac{1}{2^{i-1}}$.
Therefore, it is possible to place $2^{i-1}$ of them next to each other in one bin.
As we are allowed to slice these items horizontally, we can pack them into a container of width $1$ and depth $\frac{\eps h(G_i)}{2} \cdot \frac{1}{2^{i-1}} = \frac{\eps h(G_i)}{2^{i}}$.
Note that the total area of the items in group $G_i$  is at least $\frac{\eps h(G_i)}{2^{i}}$.
Therefore, the total depth of all these containers for groups $G_i$ is bounded by 
\[\sum_{i = 1}^{\lceil\log(1/\delta)\rceil} \frac{\eps \cdot h(G_i)}{2^{i}} \leq \eps \sum_{i = 1}^{\lceil\log(1/\delta)\rceil} \frac{h(G_i)}{2^{i}}\leq \eps \cdot \area(I) \leq \eps \cdot \opt_{\twobp}.\]

As we are allowed to slice the horizontal items horizontally, we can fill each bin to the top and therefore introduce at most $\lceil \eps \opt_{\twobp}\rceil$ new bins.
\end{proof}

\paragraph*{Rounding \bigy items}
Rounding the \bigy items presents a bigger challenge compared to rounding the horizontal or vertical items.
To round the \bigy items, the entire packing has to be rearranged to allow one side of the items to be rounded to a multiple of $\delta^2/2$.
The main ingredient for the structural result for \twobp~is the following restructuring theorem, which states that each bin can be restructured to have one of two properties, while not increasing the number of used bins too much.

\begin{property}
\label{prop1}
The width and the x-coordinate of each item in the bin of width at least $\delta$ is a multiple of $\delta^2/2$.
\end{property}

\begin{property}
\label{prop2}
The depth and the y-coordinate of each item in the bin of depth at least $\delta$ is a multiple of $\delta^2/2$.
\end{property}
\begin{theorem}[\cite{jansen2016new}]
\label{thm:2d-bin-packing-repacking-structure}
    For any $\delta$, with $1/\delta$ being a multiple of 24, and for any solution that fits into $m$ bins, we are able to round up the widths or the depths of the rectangles so that they ﬁt into $(3/2 +5\eps)m +37$ bins, while the packing of each of the bins satisfies either \cref{prop1} or \cref{prop2}.
\end{theorem}

\cref{prop1} and \cref{prop2} allow one side of the items to be rounded to a multiple of $\delta^2$. The main idea behind rounding the other side is to use linear grouping for each rounded width and each rounded depth.

\begin{lemma}
\label{lem:rounding-large-items-two-bp}
Consider a packing into $B$ bins where each bin fulfills \cref{prop1} or \cref{prop2}.
By adding at most 
$O(\eps) \opt_{\twobp}(I) +O(1)$
bins, the sizes of the \bigy items can be rounded to at most $O(1/(\delta^3\eps))$ sizes.
\end{lemma}

\begin{proof}
    First, the \bigy items are partitioned into groups. 
    Let $L_i^w$ denote that set of \bigy items where the width has been rounded to $i \delta^2/2$ for some $i \in \mathbb{N}$, where $i < 2/\delta^2$.
    Similarly, let $L_i^d$ denote the set of \bigy items where the depth has been rounded to $i \delta^2/2$ for some $i \in \mathbb{N}$, where $i < 2/\delta^2$.
    In the next step, within each group $L_i^w$, the depths are rounded using linear grouping, while within each group $L_i^d$, the widths are rounded using linear grouping.

    Consider the set $L_i^d$ for some $i \in \mathbb{N}$. 
    Sort set $L_i^d$ by width and partition it onto $\lceil 2/(\eps\delta) \rceil$ groups $L_{i,1}^h,\dots, L_{i,\lceil 2/(\eps\delta) \rceil}^h$, such that each group contains $\lfloor\eps\delta|L_i^d|/2 \rfloor$ items, except for the last group $L_{i,\lceil 2/\delta^2 \rceil}^h$ containing the items with the smallest widths. 
    This group can contain fewer items. 
    In each group, the widths are rounded to the size of the largest width in that group.
    Note that in the given packing, for each group, the rounded items can be placed in place of the original items in the group with the next larger width.
    The last group of each has $L_i^d$ will be placed into new bins.
    Note that as there are at most $\lfloor\eps\delta|L_i^d|/2 \rfloor$ items in the set, and each has a width of at most $1$ and a depth of $i \delta^2/2$ it holds that
    \[\area(L_{i,1}^h) \leq \lfloor\eps\delta|L_i^d| \rfloor \cdot i \delta^2/2\leq \eps\cdot \area(L_i^d).\] 
    Note that the last inequality holds as each item in $L_i^d$ has a width of at least $\delta$.
    
    We get a similar result when rounding the items in $L_i^w$.
    As a consequence, the total area of the items that have to be placed into additional bins is bounded by $\eps \cdot \area(L)$, where $L$ is the set of \bigy items. 
    We use Steinberg's algorithm \cite{steinberg1997strip} to place the items into a box with width $1$ and depth $2\eps\cdot \area(L)$. This box is then cut at every integer depth to create bins from the box. The items cut by the line will be placed into extra bins. Therefore, we can place these items into at most $\lceil 4 \eps \area (L) \rceil \leq \lceil4 \eps \opt_{\twobp}(I)\rceil$ bins.
\end{proof}

Note that when just considering the items, we do not know which of them are rounded to have widths that are multiples of $\delta^2$ and which of them are rounded to have depths that are multiples of $\delta^2$. 
In~\cite{jansen2016new}, it is shown that there is only a constant number of possible rounding for \bigy items, and one of them corresponds to the one derived by the method in \cref{lem:rounding-large-items-two-bp}.

\begin{lemma}[\cite{jansen2016new}]
\label{lem:iterate-roundings-two-bp}
There is an algorithm that iterates at most $n^{{O}_{\eps}(1)}$ different roundings for the \bigy items, where one of them is guaranteed to be the same, that can be derived from the optimal solution.
\end{lemma}

The main idea to find this rounding is to iterate over the possible subsets of items that are the $O(1/\delta^4)$ items responsible for the rounded width or depth in the linear grouping step. 
Further, for each rounded depth and width that is a multiple of $\delta^2$, the total number of items is guessed that are rounded this way.
After this guessing step, the \bigy items are assigned to the rounded classes using a flow network.
As there are only $n^{{O}_{\eps}(1)}$ possible choices, the algorithm needs to iterate over at most this many number of different roundings.

\begin{proof}[Proof of \cref{lem:twobp-rounding-main}]
    By \cref{lem:rounding-vert-hor-items-2bp}, we can round the depths for vertical items and the widths for horizontal items to $O((1/\eps)\log(1/\delta))$ different sizes while increasing the number of used bins by only $O(\eps \opt_{\twobp}) +O(1)$.
    Further, by combining \cref{thm:2d-bin-packing-repacking-structure} and \cref{lem:rounding-large-items-two-bp}, we know that the \bigy items can be rounded to at most $O(1/\delta^4)$ types while increasing the number of bins by at most $(\frac{1}{2} + O(\eps))\opt_{\twobp} + O(1)$.
    This proves that the items can be rounded as described in the Lemma statement while maintaining $\opt_{\twobp}^{\mathrm{slice}}(\tilde{I}) \leq ((3/2)+O(\eps))\opt_{\twobp}(I) + O(1)$.
    Finally, \cref{lem:iterate-roundings-two-bp} guarantees that the right rounding can be computed in polynomial time.
\end{proof}

\subsection{Existence of a container-based packing for 2D-BP}

We define a \emph{preliminary container-based packing} as follows.
Each bin is partitioned into at most $O(1/\delta^3)$ rectangular regions called containers. 
Further vertical items may be sliced vertically, horizontal items may be sliced horizontally, and \tin items may be sliced in both directions.
The containers have the following properties:
\begin{itemize}
        \item There are at most $1/\delta^2$ container containing only one \bigy item $i$ and having size of $i$. 
        \item There are at most $3/\delta^3$ containers that contain only horizontal and \tin items. 
        \item There are at most $3/\delta^3$ containers that contain only vertical and \tin items. 
        \item The container contains only \inter items and spans the complete bin.
\end{itemize}

\begin{lemma}
\label{lem:structure-2bp-1}
Consider a packing of items $I$ into $B$ bins, where each bin satisfies either \cref{prop1} or \cref{prop2}.  
By adding at most ${O}(\eps)B+O(1)$ bins, the packing can be transformed into a preliminary container-based packing.
\end{lemma}

\begin{proof}
In the first step, we remove all the \inter items from the bin. 
Note that these items have a total area of at most $\eps  \opt_{\twobp}(I)$.
We use Steinberg's algorithm \cite{steinberg1997strip} to place the items into a bin of width $1$ and depth at most $2\eps \opt_{\twobp}$.
This box is then partitioned into $\lceil 2\eps \opt_{\twobp} \rceil$ boxes of depth $1$. The items on each cut are placed into an extra bin. Therefore, the total number of bins added is bounded by $2\lceil 2\eps \opt_{\twobp}(I) \rceil \leq {O}(\eps)B+O(1)$. 

Next, we describe how to partition the bins into containers. 
Consider a bin that satisfies \cref{prop1}, i.e., the width and the $x$-coordinate of each item with width at least $\delta$ has a width that is a multiple of $\delta^2/2$.
First, we introduce containers for the \bigy items. 
Each \bigy item gets its own container and it is positioned exactly where the \bigy item is placed.
As each bin can contain at most $1/\delta^2$ of these items, the total number of these containers is also bounded by $1/\delta^2$.

Next, we consider all vertical strips of width $\delta^2/2$ from left to right.
For each of these strips, we create at most $1/\delta$ containers for vertical items, with their left and right borders defined by the strip’s boundaries.
To introduce the depth-aligned borders parallel to the x-axis, we iterate the packing from depth $0$ to depth $1$.
At the smallest depth where a vertical item is first encountered, we open a new container by placing a border parallel to the x-axis at this depth.
The other border of this container is then set parallel to the x-axis at the next smallest depth where the scan encounters a \bigy or a horizontal item.
Once a \bigy or horizontal item is encountered, the scan continues until either depth $1$ is reached or another vertical item appears.
In the latter case, a new container is opened, and the process is repeated.
This method ensures that no horizontal item is sliced, and vertical items are only sliced vertically along box borders, as all \bigy and horizontal items are aligned to the $\delta^2/2$ grid with their width.
As vertical items have a depth of at least $\delta$, the total number of containers for vertical items created this way is bounded by $\frac{1}{\delta} \cdot \frac{2}{\delta^2} = \frac{2}{\delta^3}$.

Finally, we create containers for the horizontal items. 
First, note that at this point, all vertical and \bigy items are covered by containers.
Therefore, we only need to partition the residual area into rectangular containers.
To do that, we consider the $y$-coordinates of all boxes of containers for vertical and \bigy items. 
We expand these upper and lower borders to the left and right until they meet another container.
These horizontal lines partition the residual area into rectangular containers.
As they do not intersect containers for vertical or \bigy items, no vertical item is sliced horizontally by these containers.
Note that each of these horizontal lines is adjacent to at most $3$ containers for horizontal items. 
As the total number of these lines is bounded by $4/\delta^3$ and each container requires two lines (upper and lower border) to be complete, there are at most $6/\delta^3$ containers for horizontal items.

Note that this partition can be done analogously to any container that fulfills \cref{prop2} by changing the role of vertical and horizontal containers.
\end{proof}

In the next step, the containers for vertical and horizontal items are divided into sub-containers, such that they only contain one type of item. 
This will reduce the total number of possible container configurations further.

Recall the definition of a $k$-\emph{2D-container-packing}:
Each bin of the packing is partitioned into containers. 
The containers are of five types, and only specific types of items from $\tilde{I}$ are allowed to be packed in the corresponding containers:
\begin{itemize}
\item (i) {\em Big container:} Each such container contains only one (rounded up) \bigy item and has the size of this \bigy item.  
\item (ii) {\em Horizontal containers:} each such container has a width $w \in \mathcal{W}$ and a depth that is a multiple of $\mu$, and contains only horizontal items with width $w$.  
The total width of these containers per bin is bounded by $O(k)$.
\item (iii) {\em Vertical containers:} 
each such container has a depth $d \in \mathcal{D}$, a height that is a multiple of $\mu$, and contains only vertical items with depth $d$.
The total depth of these containers per bin is bounded by $O(k)$.
\item (iv) {\em Tiny containers:} contain only \tin items and have a width and depth that is a multiple of $\mu$. Each bin contains at most $O(k)$ of these containers.
\item (v) {\em Intermediate containers:} These containers contain only \inter items. There will be an extra $O(\eps)\opt_{\twobp}$ bins reserved separately to pack the \inter items.
\end{itemize}

In the following, we prove that such a $k$-\emph{2D-container-packing} can be generated from a preliminary container-based packing. 
Further, note that \cref{prop1} implicitly states that each item that has a width larger than $\delta$ is further rounded up to be a multiple of $\delta^2/2$ if property. 
We reverse this rounding to multiples of $\delta^2/2$ for the horizontal items and vertical items (for \cref{prop2}). 
Instead, for them, we consider the rounded width from the first rounding step in \cref{lem:rounding-vert-hor-items-2bp}.

\begin{lemma}
Consider a preliminary container-based packing into $B$ bins.
By adding at most ${O}(\max\{\eps,\mu/\delta^3\}\cdot B)$ bins, the packing can be transformed into a $(1/\delta^3)$-2D-container-packing for $\tilde{I}$.
\end{lemma}

\begin{proof}
Assume we are given a preliminary container-based packing. 
The containers for the \bigy items already have the desired properties.
Therefore, this proof focuses on partitioning the other containers into smaller containers for vertical, horizontal, and \tin items.

\textbf{Horizontal Containers.}
Consider the containers for horizontal items. First, we round down the depths of these containers to multiples of $\mu$. 
As there are only ${O}(1/\delta^3)$ containers per bin,
rounding down the depth of each container for horizontal items in this bin to the next multiple of $\mu$ removes a total depth of at most ${O}(\mu/\delta^3)$ from all containers of this bin combined.
In this process, we slice some of the horizontal items horizontally.
The total depth of the removed parts over all bins adds up to at most ${O}(\mu/\delta^3) B$.
We create for these removed parts ${O}(\mu/\delta^3) B\geq 1$ new containers that each fill a complete bin.

In the next step, we round the widths of these containers using geometric grouping.
We round the widths to $\frac{2}{\eps} \cdot \log(\frac{1}{\delta})$ different sizes, using the same process as to round the horizontal items.
Due to the rounding, we need to place some of the containers into new bins. 
Note that the total area of the containers that need to be placed into new bins is bounded by $O(\eps)B$, as they have a total area of at most $B$.
As a consequence, we introduce at most $O(\eps)B\geq 1$ new bins that each are completely filled by a container for horizontal items. 
Denote by $K_H$ the set of rounded containers.

To partition the containers $K_H$ into sub-containers for the 2D-container-based packing, we use a linear program. 
Recall that $\mathcal{W}$ denotes the set of rounded widths for horizontal items. 
Let $\mathcal{W}_{K}$ denote the set of rounded widths of the containers $K_H$.
For each $w \in \mathcal{W}$, denote by $d_{H,w}$ the total depth of horizontal items with rounded width $w$ and for each $w' \in \mathcal{W}_K$ denote by $d_{K,w'}$ the total depth of containers with rounded width $w'$.
Further, for each $w' \in \mathcal{W}_K$, a {\em horizontal arrangement} $C_H$ denotes a multiset of rounded horizontal item widths from $\mathcal{W}$ that adds up to at most $w'$, i.e., these items fit next to each other into a container of width $w'$.
Let  $\mathcal{C}_{w'}$ be the set of all corresponding arrangements. 
For a given arrangement $C_H \in \mathcal{C}_{w'}$ and rounded horizontal item width $w \in \mathcal{W}$, we denote by $C_H(w)$ the multiplicity of width $w$ in the multiset $C_H$.
The variable $x_{C_H}$ represents the depth of arrangement $C_H \in \mathcal{C}_{w'}$. 
The configuration LP for horizontal items has the following form: 
\begin{align*}
    \sum_{C_H \in C_{w'}} x_{C_H} & = d_{K,w'} &\forall w' \in \mathcal{W}_K\\
    \sum_{w' \in \mathcal{W}_K}\sum_{C_H \in \mathcal{C}_{w'}} x_{C_H} \cdot C_H(w) &= d_{H,w} &\forall w \in \mathcal{W}\\
    x_{C_H} &\geq 0 &\forall C_H \in \bigcup_{w' \in \mathcal{W}_K} \mathcal{C}_{w'}
\end{align*}
Note that a solution to the LP exists, as the transformed packing into the containers $K_H$ can be sliced into horizontal strips at the borders (that are parallel to the x-axis) of all horizontal item slices in the containers. 
The items contained in a strip represent one (horizontal)  arrangement $C_H$ for the (horizontal) container. 
If we set the depth of this arrangement $x_{C_H}$ to the sum of the depths of all strips containing exactly this set of items in containers with the same rounded depth, we get a solution to the above LP.
A basic solution to the LP has at most $|\mathcal{W}| + |\mathcal{W}_K| \le O(\log(1/\delta)/\eps) \le O(1/\delta^2)$ non-zero components and can be found in ${O}_{\eps}(1)$ time.

We use the solution to the LP to fill the containers for horizontal items.
First, we increase the depth of each arrangement $x_{C_H}$ to the next larger multiple of $\mu$ and introduce for each arrangement a new container of depth $\mu$ that will be placed into an extra bin.
As there are at most $O(\log(1/\delta)/\eps)$ arrangements, the total depth of added horizontal containers is bounded by $O(\mu \log(1/\delta)/\eps)\leq O(\delta^2)$.
We iterate over the arrangements and fill them into the containers until they reach the depth of the container or are completely packed.
If one container is full, we continue to fill the next one. If the arrangement is completely packed, but the container is not full, the container is split horizontally at that point, and we continue to fill the next fitting arrangement into this container.
As there are at most $O(\frac{1}{\delta^2})$ arrangements, we add at most $O(\frac{1}{\delta^2})$ new containers for horizontal items by this step.
Note that the total number of horizontal containers per bin is still bounded by $O(1/\delta^3)$ and hence their total width is bounded by $O(1/\delta^3)$.

In the next step, we partition each horizontal container into at most $1/\delta$ sub-containers.
Consider a container with width $w' \in \mathcal{W}_K$ and depth $d$ that is filled with the arrangement $C_H$.
As the arrangements specify a set of horizontal items that can be placed next to each other inside the container, we partition the container vertically into sub-containers, each of depth $d$, such that, for each width $w \in C_H$, there is one sub-container that has width $w$.
We arrange the sub-containers such that they are placed next to each other.
Not that after this step, the total width of sub-containers for horizontal items is still bounded by $O(1/\delta^3)$.

Note that the widths in the arrangement might add up to something smaller than the width of the original container. 
This leaves a free space where a container for \tin items can be placed. 
This container will get the depth of the container, which is a multiple of $\mu$, and a width that is the largest multiple of $\mu$ that still fits next to the sub-containers for horizontal items.
This step introduces at most $O(1/\delta^3)$ container for small items.

\textbf{Vertical Containers.}
Note that we can partition the containers for vertical items into sub-containers similarly as for the horizontal items by just swapping width and depth. 
This creates $O_{\eps}(1)$ (sub-)containers for vertical items that have a total depth of at most $O(1/\delta^3)$, have a width that is a multiple of $\mu$ and a height $d\in \mathcal{D}$.

\textbf{Tiny Containers.} 
As mentioned before, in each container for horizontal items and each container for vertical items, we introduce a container for \tin items with width and depth that is a multiple of $\mu$.
Hence, their number is bounded by $O(1/\delta^3)$.
Note that the totally free area inside the containers for horizontal and vertical items is at least as large as the total area of the \tin items because the LP solution covers exactly the area of the horizontal (and vertical) items.
Due to the rounding of the width of sub-containers for \tin items in the containers for horizontal items, we lose an area of at most $O(\mu/\delta^3)$ per bin, as there are at most $O(1/\delta^3)$ container for horizontal items per bin with a total depth of at most $O(1/\delta^3)$.
Similarly, in containers for vertical items, we lose a total area of $O(\mu/\delta^3)$ for \tin items.
Therefore, a total area of at most $O(\mu/\delta^3) B$ of the \tin items that were contained in the containers for horizontal items cannot be covered by the newly introduced containers for \tin items. 
Therefore, we can use $O(\mu/\delta^3) B$ new bins to cover the residual area for \tin items. 
\end{proof}

To find a $k$-\emph{2D-container-packing} for all items, it is a useful tool to have an algorithm that can decide for a set of items if these items can be placed into one bin, that fulfills the requirements to be a bin of a $k$-\emph{2D-container-packing}.
In the next step, we prove that such a decision can indeed be made in polynomial time.

\begin{lemma}
\label{lem:check-if-one-bin-packing}
    Given a set of rounded items $I$, it is possible to decide in $O(n)+ O_{\delta,\mu,k}(1)$ time if this set of items can be placed into a \emph{$k$-2D-container-packing} that consists of one bin. 
\end{lemma}
\begin{proof}
First, note that for each rounded horizontal item width $w \in \mathcal{W}$, the total height of items of this width has to be bounded by $k$. 
Otherwise, the items have a total area larger than $1$ and trivially do not fit into one bin.
Furthermore, the number of containers for horizontal items is bounded by $k/\delta$ as each container has a width of at least $\delta$.
As all containers for horizontal items have a depth that is a multiple of $\mu$ and there are at most $k/\delta$ containers, there are only $O_{\delta,\mu,k}(1)$ possible choices for all the container heights of containers covering the horizontal items with width $w$.
Similarly, for each vertical item depth $\in \mathcal{D}$, there are only $O_{\delta,\mu,k}(1)$ possibilities to cover these items with containers.
Further, the containers for \tin items have depths and widths that are multiples of $\mu$, and per bin, there exist at most $O(k)$ of them.
Therefore, there are only $O_{\delta,\mu,k}(1)$ possibilities to choose container sizes for \tin items.

Given a choice for container sizes for \bigy, horizontal, vertical, and \tin items, we can check whether these containers can be placed into one bin in $O_{\delta,\mu,k}(1)$ time, as there are at most $O_{\delta,\mu,k}(1)$ container.
\end{proof}

\subsection{Algorithm to find a container-based packing for 2D-BP}

A {\em 2D bin configuration} $C \in \mathbb{N}^{|\mathcal{T}|+|\mathcal{D}|+|\mathcal{W}|+1}$ is a vector that specifies a set of items that can be placed into one bin of a 2D-container-packing. 
This vector specifies for each \bigy item type $t \in \mathcal{T}$, the number $n(t,C)$ of items of this type in the container; for each horizontal item width $w \in \mathcal{W}$,  the number $D(w,C)$ where the total depth of containers for these items is $D(w,C) \cdot \delta^4$; for each vertical item depth $d \in \mathcal{D}$, the number $V(d,C)$ where the total width of containers for these items is $V(d,C)\cdot \delta^4$; and for \tin items, the number $S(C)$  where the total area of containers for \tin items is $S(C)\cdot \delta^8$.
Let $\mathcal{C}$ be the set of all 2D bin configurations.
Then $|\mathcal{C}|$ is $O_{\eps}(1)$ as the sum of entries has an upper bound of $O(1/\delta^8)$. 
By \cref{lem:check-if-one-bin-packing}, we can verify in polynomial time if a given vector translates to a configuration $C$.

To find a 2D-container-packing, the algorithm from~\cite{jansen2016new} utilizes an integer program, to find $\opt_{\twobp}^{C}(\tilde{I})$. 
However, for our purposes, the linear programming (LP) relaxation of this program is sufficient.
For each \bigy item type $t \in \mathcal{T}$, let $n_t$ denote the number of items with this type. 
Similarly, let $W_d$ denote the total width of vertical items with rounded depth $d \in \mathcal{D}$, 
and $D_w$ denote the total depth of horizontal items with rounded width $w \in \mathcal{W}$. 
Further, 
let $S_{\area}$ denote the total area of \tin items in $I$.
We introduce variables $x_C$ that denote for each configuration $C \in \mathcal{C}$, the {\em amount} of this configuration in the solution, e.g., if $x_C= 2.7$, the configuration $C$ appears 2.7 times in the optimal LP solution. 
The LP is defined as follows. 

\begin{align*}
    \min \sum_{C \in \mathcal{C}} x_C\\
    \text{s.t.} \quad \quad \sum_{C \in \mathcal{C}} n(t,C)x_C &\geq  n_t &\forall t \in \mathcal{T}\\
     \sum_{C \in \mathcal{C}} V(d,C)\delta^4x_C &\geq  W_d &\forall d \in \mathcal{D}\\
    \sum_{C \in \mathcal{C}} D(w,C)\delta^4x_C &\geq  D_w &\forall w \in \mathcal{W}\\
    \sum_{C \in \mathcal{C}} S(C)\delta^8x_C &\geq  S_{\area}\\
    x_C &\geq 0 &\forall C \in \mathcal{C}
\end{align*}

Let $\opt_{\twobp}^{\text{LP}}(\tilde{I})$ denote the optimal solution for the LP. 
A basic solution to this LP uses only $O_{\eps}(1)$ different configurations, as there are only $O_{\eps}(1)$  types of containers in total.

\begin{restatable}{lemma}{LPVsContainerTwobp}
\label{lem::LP-vs-container-twobp}
$\opt_{\twobp}^{\text{LP}}(\tilde{I}) \leq \opt_{\twobp}^{C}(\tilde{I})$ and in $n^{O_{\eps}(1)}$ time a basic solution $x$ with  $\sum_{C \in \mathcal{C}} x_C = \opt_{\twobp}^{\text{LP}}(\tilde{I})$ can be found.
\end{restatable}

\begin{proof}
We show that $\opt_{\twobp}^{LP}(\tilde{I}) \leq \opt_{\twobp}^{C}(\tilde{I})$ and in $O_{\eps}(1)$ time a basic solution $x$ with  $\sum_{C \in \mathcal{C}} x_C = \opt_{\twobp}^{LP}(\tilde{I})$ can be found.
    Each 2D-container-packing can be transformed to be a solution to the LP by creating the corresponding configuration for each bin and adding one to the entry of $x$ corresponding to the constructed configuration, implying  $\opt_{\twobp}^{LP}(\tilde{I}) \leq \opt_{\twobp}^{C}(\tilde{I})$.
    The number of variables is in $O_{\eps}(1)$ and the number of constraints is in $O_{\eps}(1)$. Therefore, a basic optimal solution can be found in $O_{\eps}(1)$ time, e.g., even by using the simplex algorithm.
\end{proof}

Now we have all the tools to prove \cref{thm:finding-2D-container-packing}.
In the proof, we discuss the algorithm that 
given some instance $I$ for \twobp, finds in polynomial time a rounded instance $\tilde{I}$ with $|\mathcal{T}|, |\mathcal{D}|, |\mathcal{W}| \in O(1/\delta^4)$ and a $(1/\delta^3)$-2D-container-packing of $\tilde{I}$ into $B$ bins with at most $O_{\eps}(1)$ different bin-configurations that fulfills $B \leq (\frac{3}{2}+O(\eps))\opt_{\twobp}(I) +O_{\eps}(1)$.


\begin{proof}[Proof of \cref{thm:finding-2D-container-packing}]
The basic idea of the algorithm is to consider each of the $O_{\eps}(1)$ possible roundings to create $\tilde{I}$ described in \cref{lem:twobp-rounding-main}.
For each of these $\tilde{I}$, the algorithm finds the optimal solution to the configuration LP. 
Note that the set of all feasible configurations can be found in polynomial time using \cref{lem:check-if-one-bin-packing}.
For each solution, $x$ to the LP, the algorithm creates $\lceil x_C\rceil$ bins for each $C \in \mathcal{C}_{\tilde{I}}$.
Note that the set of possible configurations differs for different roundings $\tilde{I}$.
For each of these $\lceil x_C\rceil$ bins, the algorithm creates the same $(1/\delta^3)$-2D-container packing by using the algorithm from  \cref{lem:check-if-one-bin-packing} to create a 2D-container-packing.
Finally, it returns the $(1/\delta^3)$-2D-container packing with the smallest number of bins over all the different roundings for $I$.
Let $B$ be the smallest number of required bins.

First, note that the number of configurations is bounded by $O_{\eps}(1)$, as a basic optimal solution to the configuration-LP has at most $O_{\eps}(1)$ non-zero components.
Further, by \cref{lem::LP-vs-container-twobp} it holds that $\opt_{\twobp}^{\text{LP}}(\tilde{I}) \leq \opt_{\twobp}^{C}(\tilde{I})$ where $\opt_{\twobp}^{C}(\tilde{I})$ refers to an optimal $(1/\delta^3)$-2D-container packing for $\tilde{I}$.
Further, \cref{lem:twobp-rounding-main} states that for one of the iterated rounded instances $\tilde{I}$ it holds that $\opt_{\twobp}^{C}(\tilde{I}) \leq (\frac32+{O}(\eps))\opt_{\twobp}(I) + {O}_{\eps}(1)$.

Finally, as there are at most $O_{\eps}(1)$ different configurations in the basic LP solution, it holds that $B \leq \opt_{\twobp}^{\text{LP}}(\tilde{I}) +O_{\eps}(1)$.
Hence, in total, the solution returned by the algorithm uses at most $(\frac32+{O}(\eps))\opt_{\twobp}(I) + {O}_{\eps}(1)$ bins with at most $O_{\eps}(1)$ different configurations.
\end{proof}



\end{document}